\documentclass[twocolumn,tighten,twocolappendix]{aastex631}
\usepackage[T1]{fontenc}
\usepackage{lmodern}
\usepackage{amsmath,amssymb}
\usepackage[dvipsnames]{xcolor}
\usepackage{amsmath,amssymb}
\usepackage[dvipsnames]{xcolor}
\usepackage{graphicx}
\usepackage{multirow}
\newcommand{\tcell}[1]{ \parbox[c]{0.21\textwidth}{\centering #1}}
\newcommand{\dd}{\mathrm{d}}

\newcommand{\Msun}{M_\odot}
\newcommand{\Rsun}{R_\odot}
\newcommand{\MBH}{M_{\mathrm{BH}}}

\newcommand{\Ms}{M_\star}

\newcommand{\Rs}{R_\star}

\newcommand{\Bs}{B_\star}
\newcommand{\Phis}{\Phi_\star}
\newcommand{\rt}{R_{\rm t}}
\newcommand{\rp}{R_{\rm p}}
\newcommand{\rg}{R_{\rm g}}
\newcommand{\rh}{r_{\rm H}}
\newcommand{\LBZ}{L_{\text{BZ}}}
\newcommand{\enh}{\mathcal{E}}  
\newcommand{\ergs}{\mathrm{erg\,s^{-1}}}
\newcommand{\lta}{\lower 2pt \hbox{$\, \buildrel {\scriptstyle <}\over {\scriptstyle \sim}\,$}}
\newcommand{\gta}{\lower 2pt \hbox{$\, \buildrel {\scriptstyle >}\over {\scriptstyle \sim}\,$}}
\definecolor{burntorange}{RGB}{120,50,12}

\newcommand{\ulemsout}{\leavevmode \bgroup \ULdepth=-.55ex \ULset}
\makeatletter
\newcommand{\editsout@strike@citep}[1]{\mbox{\editsout@orig@citep{#1}}}
\newcommand{\editsout@strike@citet}[1]{\mbox{\editsout@orig@citet{#1}}}
\newcommand{\editsout@strike@cite}[1]{\mbox{\editsout@orig@cite{#1}}}
\newcommand{\editsout}[1]{%
  \def\editsouttemp{#1}%
  \let\editsout@save@citep\citep
  \let\editsout@save@citet\citet
  \let\editsout@save@cite\cite
  \let\citep\editsout@strike@citep
  \let\citet\editsout@strike@citet
  \let\cite\editsout@strike@cite
  \expandafter\ulemsout\expandafter{\editsouttemp}%
  \let\citep\editsout@save@citep
  \let\citet\editsout@save@citet
  \let\cite\editsout@save@cite
}
\AtBeginDocument{%
  \global\let\editsout@orig@citep\citep
  \global\let\editsout@orig@citet\citet
  \global\let\editsout@orig@cite\cite
}
\makeatother

\begin{document}
\title{The Origin of the Magnetic Flux Driving TDE Jets}
\author{Nimrod Tripto}
\affiliation{Racah Institute of Physics, The Hebrew University of Jerusalem, Israel}
\email{nimrod.tripto@mail.huji.ac.il}

\author{Julian Krolik}
\affiliation{Department of Physics and Astronomy, Johns Hopkins University, Baltimore, USA}

\author{Tsvi Piran}
\affiliation{Racah Institute of Physics, The Hebrew University of Jerusalem, Israel}
\begin{abstract}
Tidal Disruption Events (TDEs) occur when a star approaches a black hole closely enough to be torn apart by tidal forces, after which the stellar debris begins to orbit the black hole. Over the past decade, hundreds of TDEs have been identified, with many more expected from upcoming surveys. A small subset of these events launch transient, highly luminous relativistic jets ($10^{47}-10^{48}\ergs$ isotropic-equivalent). These are generally interpreted in terms of the Blandford--Znajek mechanism, implying the presence of substantial magnetic flux near the black hole horizon. The question arises: What is the origin of this flux?
In this paper, we investigate three candidate sources: 
stellar magnetic fields, magnetic flux from large radii around the black hole (the Lasso mechanism), and magnetic flux 
stored in the inner portion of a pre-existing accretion disk. We find that: 
observed stellar magnetic fields are insufficient to power these jets;  the Lasso mechanism requires
shallow radial magnetic field profiles and a mechanism to trap the magnetic flux brought by the debris in the vicinity of the black hole;
pre-existing  
magnetic flux near the black hole in an accretion disk
is the most plausible source.
\end{abstract}

\section{Introduction}
\label{sec:introduction}
In most TDEs, there is little evidence that the light we can see was generated in a jet: most have spectra best-described in terms of thermal radiation both in the optical/ultraviolet and X-ray bands \citep{Gezari2021,Guolo+2024}. However, of the hundreds of events cataloged as likely TDEs, four \citep{Bloom+2011,Burrows+2011,Cenko+2012,Brown+2015,Andreoni+2022,Pasham+2023} display the hallmarks of jet radiation: nominal luminosity well above Eddington, with the bulk of the radiation in hard X-rays having a power-law spectrum.
Rapid, large-amplitude X-ray variability provides an additional diagnostic for these events, as such extreme variability on short timescales is less typical of standard AGN emission, and is commonly interpreted as further evidence for a relativistic jet \citep{Bloom+2011,Burrows+2011}.
With isotropic X-ray luminosities in the range of $10^{47} -10^{48}\,\ergs$, the actual jet luminosity after allowance for beaming is still quite high (e.g., $10^{44} - 10^{45}\,\ergs$), and the Poynting flux power of the jet must be even greater than the actual jet luminosity.

Relativistic jets in accreting black hole systems are widely understood to be powered by the Blandford--Znajek (BZ) mechanism. This process has two prerequisites: high black-hole spin, which supplies the energy reservoir, and sufficiently large-scale poloidal magnetic flux near the horizon, which enables that energy to be extracted as outgoing Poynting flux. 
However, simple estimates show that the magnetic flux advected by a solar-like star undergoing tidal disruption falls short, by several orders of magnitude, of the flux required to power the observed luminosities of jetted TDEs. This discrepancy poses a fundamental problem: if an ordinary disrupted star cannot supply sufficient magnetic flux on its own, what is the origin of the large-scale magnetic field required to launch and sustain a powerful relativistic jet?
Addressing this question is essential for understanding the conditions under which TDEs can produce jets, and motivates the exploration of alternative sources and transport mechanisms for magnetic flux in these systems.

We consider in detail three candidate sources for the supply of magnetic flux, beginning with the two that have received the most attention in the literature.
First, we examine the simplest possibility, in which the flux originates from the star itself, i.e., the pre-existing stellar magnetic field carried by the disrupted debris. 
Although the solar-like flux is insufficient, stellar magnetic flux can vary substantially between different stellar types. 
We therefore consider a range of possible progenitors, focusing on those that might carry unusually large magnetic fluxes.

We then consider the Lasso mechanism, in which magnetic flux, perhaps associated with a dormant accretion flow, is transported by the TDE debris stream from large radii to the vicinity of the black hole  \citep{TchekhovskoyMetzger2014,KelleyTchekhovskoy2014}. 

In our third candidate scenario,
a TDE occurs close to a supermassive black hole surrounded by
a pre-existing accretion flow,
and the impact of the debris stream on that accretion disk causes the disk matter within the disrupted star's pericenter to be swept inward \citep{Chan2019}.  Here we suppose that the disk's magnetic flux is likewise transported to the black hole's horizon.
For the last two models, consistency with observations demands that there are no powerful pre-existing jets or other sources of emission from the galactic nucleus. 

Another model discussed in the literature, which we do not examine in detail here, is that the magnetic flux powering the jet is generated within a compact accretion flow formed after the disruption \citep{KrolikPiran2012,Piran2015SadTch,Dai+2018,CurdNarayan2019}. This scenario requires the debris to circularize and form an inner disk, a process that may be delayed by the time needed to dissipate the orbital energy \citep{Piran+2015,Krolik+2025}. Such a delay is difficult to reconcile with observed jetted TDEs, where the jet appears to be launched promptly. Although late-time outflows have been suggested, most notably in AT2018hyz, whose radio emission was interpreted as evidence for an outflow launched $\sim700$ days after disruption \citep{Cendes+2022}, this interpretation is not unique, and the event has also been modeled as an off-axis jet launched promptly \citep{Sfaradi+2024}. We therefore do not examine this mechanism further.

We begin by describing the astrophysical setup and our conventions in Section~\ref{sec:physicalsetup}. We discuss intrinsic stellar magnetic fluxes in Section~\ref{sec:stellarflux}. We analyze flux supply from large radii (the Lasso mechanism) in Section~\ref{sec:Lasso}.  We study flux supply from accretion-disk disturbance in Section~\ref{sec:TDEinAcc}. 
We discuss the implications of these results in Section ~\ref{sec:discussion}.
We summarize the main conclusions in Section~\ref{sec:summary}.

\section{Astrophysical Background}
\label{sec:physicalsetup}
\subsection{The Blandford-Znajek mechanism}
\label{sec:BZmechanism}
The success of the Blandford-Znajek (BZ) mechanism \citep{BZ1977} in readily generating jets when rotating black holes accrete poloidal magnetic field in simulations \citep{McK+Gammie2004,HK2006,BHK2008,BHK2009,Tchekhovskoy2011} and in explaining specific features in jet images \citep{EHTV2019,EHT_M87_Magnetic_Field_2021} strongly suggests that this mechanism explains TDE jets as well.

To quantify the magnetic-flux requirement as described in Section~\ref{sec:introduction}, we begin from the standard BZ scaling for jet power, written in the convention of \citet{Tchekhovskoy2011} and \citet{KelleyTchekhovskoy2014} as\footnote{In \citet{KelleyTchekhovskoy2014}, the expression used is $\LBZ = c \xi_{\rm BZ}\Phi^2/2\pi^2 R_{\rm s}^2$.
We substitute $\rg$ for $R_s$ as it is more appropriate for spinning black holes, and is more numerically consistent with our choice of $\xi_{\rm BZ}$.}
\begin{equation}
\LBZ = \frac{c \xi_{\rm BZ}}{2\pi^2 \rg^2} \Phi^2,
\label{eq:LBZ}
\end{equation}
where $\rg \equiv GM_{\rm BH}/c^2$ is the gravitational radius of a black hole of mass $\MBH$ and $\Phi$ is the magnetic flux 
threading the black hole horizon.
The efficiency $\xi_{\rm BZ}$ encodes the spin-dependence of the order-unity factor required by what is fundamentally an expression derived from dimensional analysis. Its exact form is unsettled \citep{McKinney2005,HK2006,Tchekhovskoy2011}, but all agree that it is a strongly increasing function of the spin parameter. 
We adopt $\xi_{\rm BZ}\approx0.1$, corresponding to a high black hole spin, 
for numerical evaluations.
In the papers introducing this scaling expression,  
$\Phi$ is defined as
\begin{equation}
\label{eq:PhiBH}
\Phi = \frac{1}{2} \int_{\text{horizon}} \sqrt{-g} d\Omega |B_r|,
\end{equation}
where $g$ is the metric determinant.
\footnote{When substituted into Eq.~\eqref{eq:LBZ}, this definition yields a dependence on magnetic field resembling (although imperfectly) the exact form in Kerr-Schild coordinates (Poynting flux $\propto (B^r)^2$) derived by \citet{McK+Gammie2004}.  However, the goal of this paper is to estimate $L_{\rm BZ}$ when a magnetic field is brought to a black hole from far away.  For this purpose, the physical magnetic flux, which includes the field's sign and is integrated over an open surface (e.g., a single hemisphere of the horizon), is more appropriate because, unlike the ``flux'' defined by Eq.~\eqref{eq:PhiBH}, it is rigorously conserved. This distinction is, however, immaterial at the order-of-magnitude level of this paper.}
The magnetic flux required for a given  $L_{\rm BZ}$ is then easily inferred
\begin{eqnarray}
\label{eq:Phi-from-LBZ}
\Phi
&\sim&
1.2 \cdot 10^{29} ~\text{G}~\text{cm}^{2}
\left(\frac{\MBH}{10^6\Msun}\right)
\left(\frac{\xi_{\rm BZ}}{0.1}\right)^{-1/2}
\\
\nonumber
&\times&
\left(\frac{L_{\rm BZ}}{10^{44}\ergs}\right)^{1/2},
\end{eqnarray}
using $\Phi = A_H B_H/2=4\pi \rg^2 B_H \rh$ ($A_H$ taken from \citet{Hobson+2007}), where $\rh=R_H/\rg=1+\sqrt{1-a_*^2}$ ($a_*$ the spin of the black hole) is the horizon radius and $B_H$ is the mean radial field on the horizon. We can then find the field required:
\begin{eqnarray}
\label{eq:B-from-LBZ}
B_H
&\sim&
4.3 \cdot 10^{5} ~\mathrm{G} \,
\left(\frac{\MBH}{10^6\Msun}\right)^{-1}
\left(\frac{\xi_{\rm BZ}}{0.1}\right)^{-1/2}
\\
\nonumber
&\times&
\left(\frac{L_{\rm BZ}}{10^{44}\ergs}\right)^{1/2}
\rh^{-1}.
\end{eqnarray}
Here we have scaled to a Poynting flux power $10^{44}\,\ergs$, the amount associated with an isotropic jet power equal to $\sim10^{47}\,\ergs$ and a beaming-correction factor of $\sim10^{3}$.  

For a proposed mechanism to be at all viable, it must, of course, be capable of supplying enough magnetic flux to power the jets we see.  In addition, for the latter two schemes mentioned above, the luminosity from any pre-existing jet or accretion disk must lie below observational upper bounds. 
Another criterion important for the latter two is that the mechanism should act on a timescale comparable to or shorter than the flare duration, typically $\sim 1$~month.

We put aside for the time being subsidiary considerations such as the greater jet-powering efficiency of dipolar field geometries as compared to toroidal fields and higher-order multipole poloidal fields \citep{BHK2008}. 
We also ignore for the time being the potentially large ratio between jet power due to Poynting flux near the event horizon and the luminosity in photons the jet radiates at larger distances.

\subsection{Tidal disruptions}
\label{sec:TDEs}
The tidal radius $\rt$ and the return time of the debris $t_0$ 
are defined as \citep{Rees1988}
\begin{eqnarray}
\label{eq:r_t_correction}
r_t \equiv \frac{\rt}{\rg}
&\sim&
23\,
\left(\frac{\Rs}{\Rsun}\right)
\left(\frac{\MBH}{10^6\,\Msun}\right)^{-2/3}
\\
\nonumber
&\times&
\left(\frac{\Ms}{\Msun}\right)^{-1/3}
\left(\frac{\Psi}{0.5}\right),
\end{eqnarray}
where $R$ denotes values in conventional units while $r$ denotes values in units of the black hole's gravitational radius, $\rg$. 
We determine the Keplerian orbital parameters by setting the pericenter distance to $r_t$. The resulting apocenter distance is \begin{eqnarray} 
&&\hat{a}\approx \left(\frac{\Rs}{\rg}\right)\left(\frac{\MBH}{\Ms}\right)^{2/3} 
\Xi^{-1} \approx 3400~\left(\frac{\Rs}{\Rsun} \right )
\\ &&\times \nonumber 
\left(\frac {\MBH}{10^6\Msun}\right )^{-1/3}\left(\frac{\Ms}{\Msun}\right)^{-2/3}\left(\frac{\Xi}{1.47}\right)^{-1} \ .
\end{eqnarray} The corresponding semi-major and semi-minor axes are $a_0\approx \hat{a}/2$ and $b\approx (\Rs/\rg)(\MBH/\Ms)^{1/2}$, respectively.  For $\MBH\gg\Ms$, the eccentricity is $\varepsilon=1-2\Xi\Psi \,(\Ms/\MBH)^{1/3}$. 
The return time of the fastest-returning matter is
\begin{equation} 
\label{eq:t_0_correction} 
t_0 \sim 23 \, \text{d} \, \left(\frac{\MBH}{10^6 \Msun}\right)^{1/2} \left(\frac{\Rs}{\Rsun}\right)^{3/2} \left(\frac{\Ms}{\Msun}\right)^{-1} \left(\frac{\Xi}{1.47}\right)^{-3/2}. 
\end{equation}
Throughout this paper, when computing TDE parameters, we use, following \citet{Ryu+2020b, Ryu+2020a}, the correction factors for the maximum radius for full disruption and the orbital energy of the most tightly-bound material, $\Psi(\MBH,\Ms)$ and $\Xi(\MBH,\Ms)$, respectively.  These factors take into account both general-relativistic effects and realistic main-sequence density profiles within the stars.
We also use a single power-law fit to the middle-aged main-sequence mass-radius relation, $\Rs \simeq \Rsun (M_*/M_\odot)^{0.88}$ \citep{Ryu+2020b, Ryu+2020a}.

Recent simulations of TDEs around black holes without pre-existing gaseous environments \citep{ShocksPowerTDEs2023,SteinbergStone2024,Price+2024,Abolmasov2025}, summarized by \citet{Krolik+2025}, show that only a small fraction of the returning debris is accreted promptly by the black hole, certainly $\lesssim 1\%$. This differs from the older picture in which the debris was expected to circularize rapidly into a compact disk with radial scale $\sim r_p$ and accrete efficiently. This result presents a challenge for models in which the magnetic flux is supplied by the TDE stream itself, such as stellar flux (Sec.~\ref{sec:stellarflux}) or Lasso (Sec.~\ref{sec:Lasso}): if most of the debris returning to the vicinity of the black hole turns around and moves outward again rather than reaching the horizon, most of the magnetic flux carried by the stream is unlikely to be delivered to the black hole. It also raises a similar problem for the scenario discussed in Sec.~\ref{sec:introduction}, in which the magnetic flux powering the jet is generated within a compact accretion flow formed after the disruption. The promptly accreted $\lesssim 1\%$ of the stream is likely insufficient to form a disk massive enough to support the required magnetic flux. Only rare events with special stream geometries, such as the extremely relativistic TDEs explored in \citet{Ryu+2023}, may allow a larger fraction of the debris to move inward. 

\subsection{Jetted TDEs}
\label{sec:jettedTDEs}

Jetted TDEs are observationally distinct from the much larger population of optical/UV TDEs. Rather than being identified primarily through thermal emission from the disruption, they are discovered through luminous nonthermal high-energy emission and, in several cases, radio afterglows \citep{Bloom+2011,Burrows+2011,Cenko+2012,Brown+2015,Andreoni+2022,Pasham+2023}. This difference in selection is important: relativistic beaming allows jetted events to be detected to much larger distances than ordinary thermal TDEs, but the same beaming also makes their intrinsic rate difficult to infer.

The best-studied event, Swift J1644+57, is often treated as the archetype of the class, but it may not be representative of all jetted TDEs. Its inferred energetics are especially large, with a total jet energy at least $\sim 10^{51}$--$10^{52}\,\rm erg$ \citep{BarniolDuran2013, Berger2012,Zauderer2013}, 
whereas the other candidates are either less extreme or less well constrained. Radio follow-up of optically selected TDEs suggests that most TDEs are not accompanied by relativistic jets as powerful as Swift J1644+57 \citep{vanVelzen2013,Alexander2017,Alexander2020}. This fact supports the view that powerful jet production requires special conditions, rather than being a generic outcome of stellar disruption. In the context of this work, those special conditions are interpreted as conditions that allow enough large-scale magnetic flux to reach the black hole horizon.

An important observational characteristic that currently confirmed TDEs satisfy is 
the lack of prior activity at the nucleus of the host galaxy. In fact, in some surveys it is a condition for the identification of events as TDEs rather than
fluctuating AGNs. With this in mind 
we require that the luminosity of any pre-existing jet, $L^{\rm s}_{\rm jet}$, be much smaller than the TDE jet; somewhat arbitrarily, we choose a factor of 100 fainter. 
Similarly, we demand that any pre-existing accretion disk should not be more luminous than $L^{\rm max}_{\rm acc} \sim 10^{43}\,\ergs$.

\subsection{Magnetic flux transport}
\label{sec:enh}
Two of the three mechanisms discussed in this work, Lasso (Sec.~\ref{sec:Lasso}) and disturbed accretion (Sec.~\ref{sec:TDEinAcc}), share a common mathematical framework. 
In both cases, the black-hole environment is threaded by a pre-existing large-scale magnetic field, which we assume has a power-law radial intensity profile, $B(r) = B_0 r^{-q}$, where $r\equiv R/\rg$ is the dimensionless spherical radius.
Prior to the disruption, the event horizon, which is at $\rh \equiv R_H/\rg = 1 + \sqrt{1-a^2_*}$, already carries a flux corresponding to the local field,  $B_H^{\rm s}\equiv B(r=\rh) = B_0 r_H^{-q}$, and this flux powers a steady jet of
luminosity $L_{\rm jet}^{\rm s}\propto\Phi_s^2$ (Sec.~\ref{sec:BZmechanism}).

{In the Lasso model,} the TDE debris then acts as a flux collector, {whereas in the disturbed disk model, the debris triggers a sudden inflow of the inner accretion disk, and this gathers the flux}. In both cases, the disruption sweeps
up the field lines crossing some collection surface of area $S$ located
away from the hole, and advects the collected flux
$\Phi_{\rm TDE} = \langle S B \rangle
\equiv \int_S \vec{B}\cdot\dd \vec{A}
$
onto the horizon. 
The corresponding mean field at the flux's original location is $\langle B\rangle \equiv \langle S B
\rangle / S$. Here we implicitly assume that the magnetic field lines are added coherently, i.e. collected flux does not cancel. Spread over the horizon, this flux
corresponds  to a field
\begin{equation}
B_H^{\rm TDE}\sim \langle S B\rangle/(4\pi \rg^2 \rh)
= \langle B\rangle S/(4\pi\rg^2\rh) \ .
\end{equation} 
Thus, flux conservation amplifies the  field on the horizon relative to the  in situ field
by the ratio of the field-weighted collection area to the horizon area. 
The net change of the horizon
field is then set by the competition between the large area ratio and 
the considerably weaker field at larger radii from the black hole.

It is therefore natural to measure the net effect of
the TDE by a single dimensionless number, the {\it enhancement factor}
\begin{equation}
\label{eq:enh-def}
\enh
\equiv
\frac{\Phi_{\rm TDE}}{\Phi_H^{\rm s}}
=
\frac{B_H^{\rm TDE}}{B_H^{\rm s}}
=
\frac{\langle S B \rangle}{4\pi \rg^2\, \rh^{1-q}\, B_0},
\end{equation}
where $B_0$ is the pre-existing field at $R=\rg$.
Since $\LBZ\propto \Phi^2$, for $\enh+1\sim\enh$ the transient-to-steady luminosity
contrast is  given by
\begin{equation}
\label{eq:enh-ratio}
\frac{L_{\rm jet}^{\rm TDE}}{L_{\rm jet}^{\rm s}} = \enh^2 .
\end{equation}
The observational requirement that the transient jet outshine any
pre-existing jet by $\gtrsim 100$ thus translates into the compact
condition $\enh \gtrsim 10$.

Eq.~\eqref{eq:enh-def} makes transparent how a TDE can produce a
large contrast, and under what conditions. 
Consider a collector
that reaches a characteristic radius $r_c$ and covers a fraction
$f_{\rm cov}$ of the area within $r_c$. 
The flux it gathers is roughly
$
\langle S B\rangle \sim f_{\rm cov}\,\pi \rg^2 r_c^{2-q}
B_0  
$,
and therefore
\begin{equation}
\label{eq:enh-scaling}
\enh \sim \frac{1}{4}f_{\rm cov}
\left(\frac{r_c}{\rh}\right)^{2-q}\rh .
\end{equation}
This expression shows that the enhancement is not produced because the
field at large radii is strong. In fact, the mean field sampled by the
collector is much weaker than the field already present on the horizon,
$\langle B\rangle/B_H^{\rm s}\sim (r_c/\rh)^{-q}\ll 1$. Enhancement
is possible only when the collecting area grows with radius faster
than the field strength declines.

The same point can be stated in terms of the enclosed magnetic flux:
\begin{equation}
\label{eq:totalfluxlasso}
\Phi(r) \sim \rg^2 \int_{\rh}^r B_0 r'^{-q} r'\,dr'
\propto r^{2-q}.
\end{equation}
Thus, for 
$q>2$, the flux is dominated by the inner region, so
dragging in  field from large distances cannot substantially exceed the
flux already present near the black hole. We therefore restrict attention
to $q<2$ throughout this work.

Eq.~\eqref{eq:enh-scaling} also exhibits the trade-off that
distinguishes the two mechanisms: $\enh$ can be made large either by
reaching to a large collection radius at a small covering fraction, or
by collecting everything within a modest radius. The Lasso and the disturbed disk are, respectively, these two limits, and the comparison
between them carried out in Sec.~\ref{sec:Newlassovsaccretion} amounts to
comparing their enhancement factors: for a common field normalization
$(B_0,q)$,
$L_{\rm jet}^{\rm Lasso}/L_{\rm jet}^{\rm dist} =
(\enh_{\rm Lasso}/\enh_{\rm dist})^2$.

Finally, we note that $\enh$ is bounded from above independently of the
supply of flux: the horizon cannot hold an arbitrarily enhanced field.
Once $B_H^{\rm TDE}=\enh\, B_0 \rh^{-q}$ approaches the MAD saturation
value, additional collected flux is not
retained, and the effective enhancement saturates (see Sec.~\ref{sec:MAD}). 

\section{Intrinsic Stellar Magnetic Flux}
\label{sec:stellarflux}
The simplest scenario already discussed in the literature \citep{TchekhovskoyMetzger2014, Piran2015SadTch} is that the event supplies only the flux inherent to the disrupted star itself. 
The conserved magnetic flux in the star, $\Phis$, can be estimated as
\begin{equation}
    \Phis \sim \pi \Rs^2 \Bs \sim 1.5 \cdot 10^{22} \left(\frac{\Rs}{\Rsun}\right)^2 \left(\frac{\Bs}{1\hbox{~G}}\right) \hbox{~G~cm$^2$}
\end{equation}
where $\Rs$ is the stellar radius and $\Bs$ is the characteristic magnitude of the star's magnetic field.
Following the disruption, if the magnetic flux is fully advected with the stellar debris towards the black hole (which is a very favorable assumption - see Sec.~\ref{sec:TDEs}), 
the Blandford-Znajek luminosity is given by
\begin{eqnarray}
\LBZ
&\sim&
1.6 \cdot 10^{30}~{\rm erg~s^{-1}}
\left(\frac{\Bs}{1~{\rm G}}\right)^2
\label{eq:LBZ_stellar}
\\
\nonumber
&\times&
\left(\frac{\MBH}{10^6~\Msun}\right)^{-2}
\left(\frac{\xi_{\rm BZ}}{0.1}\right)
\left(\frac{\Rs}{\Rsun}\right)^4 ,
\end{eqnarray}
using Eq.~\eqref{eq:LBZ}. This is, of course, far too little to explain the observed jets. 
There are a number of stellar types that may have stronger magnetic fields,
including ordinary main-sequence stars, strongly magnetized pre-main-sequence (T-Tauri) stars, fossil-field stars such as Ap/Bp stars and rare massive O/B stars, and magnetic white dwarfs.
For each of these, we have collected typical magnetic field strengths from the literature and estimated their magnetic fluxes (see Appendix~\ref{sec:different-stars-toroidal-field}).
The results for all the stellar types we consider are summarized in Table~\ref{tab:stellar-flux-summary}. 
For all surveyed star types, the obtained BZ jet power is below the required amount by at least a factor of $\sim 10$,
and the most favorable cases are generally very rare stars.
Thus, the magnetic field measured in known classes of stars does not provide a promising explanation for the observed jetted TDE population.

\begin{table*}[ht]
\centering
\tabletypesize{\small}
\begin{tabular*}{0.9\textwidth}{@{\extracolsep{\fill}}l@{\hspace{6pt}}lcccc}
\hline\hline
& \colhead{Source} &
\colhead{$\Bs^{\rm pol}$} & \colhead{$\Rs$} &
\colhead{$L_{\rm BZ}$ (Eq.~\eqref{eq:LBZ_stellar})} &
\colhead{} \\
\hline
\multirow{5}{*}{\parbox{2.9cm}{\centering\textcolor{gray}{\small Ordinary\\stellar\\flux}}} &
Solar-like MS & $1\,{\rm G}$ & $1\,\Rsun$ & $10^{30}$~erg\,s$^{-1}$ & $\sim10^{14}$ short \\
& T-Tauri & $3\cdot10^3\,{\rm G}$&  $3\,\Rsun$ & $10^{39}$~erg\,s$^{-1}$ & $\sim10^{5}$ short \\
& Ap/Bp & $3\cdot 10^4\,{\rm G}$&  $5\,\Rsun$ & $10^{42}$~erg\,s$^{-1}$ & Rare; $\sim100$ short \\
& O/B fossil & $2\cdot 10^4\,{\rm G}$&  $10\,\Rsun$ & $6\cdot 10^{42}$~erg\,s$^{-1}$ & Best MS; $\sim20$ short \\
& Magnetic WD & $10^9\,{\rm G}$&  $10^8\,{\rm cm}$ & $7\cdot 10^{36}$~erg\,s$^{-1}$ & Area-limited; $\sim10^{7}$ short\\
\hline
\multirow{3}{*}{\parbox{2.9cm}{\centering\textcolor{gray}{\small Enhanced\\flux mechanisms}}} &
Toroidal Ap/Bp & $10^5\,{\rm G}$&  $5\,\Rsun$ & $10^{43}$~erg\,s$^{-1}$ & Rare; $\sim10$ short\\
& MS merger remnant & $2.5\cdot10^4\,{\rm G}$&  $10\,\Rsun$ & $10^{43}$~erg\,s$^{-1}$ & Rare; $\sim10$ short\\
& O/B partial-disruption remnant & $4\cdot10^5\,{\rm G}$&  $10\,\Rsun$ & $2\cdot10^{45}$~erg\,s$^{-1}$ & Very rare\\
\hline\hline
\end{tabular*}
\caption{\label{tab:stellar-flux-summary} Representative intrinsic stellar magnetic-flux estimates from Eq.~\eqref{eq:LBZ_stellar}, assuming $\xi_{\rm BZ} \sim 0.1, \MBH \sim 10^6 \Msun$ and coherent large-scale poloidal flux during disruption (toroidal Ap/Bp: effective $\Bs^{\rm tor}$ with $\Bs^{\rm tor}/\Bs^{\rm pol}\sim 2$--$5$). Representative $\Bs^{\rm pol}$ and $\Rs$ are from \citet{Stenflo2013} for solar-like main-sequence stars ($1\,\Rsun$ is fiducial), \citet{JohnsKrull2007} for T-Tauri stars, \citet{Romanyuk2025,Babcock1960_2} for Ap/Bp stars (tabulated values are Babcock's star), \citet{Braithwaite2004,Braithwaite2006} for toroidal Ap/Bp equilibria, \citet{David-Uraz2021,Wade2012_2} for magnetic O/B stars (NGC~1624-2), \citet{Schneider2019,Mandel2015} for MS merger remnants ($\Phi\sim 4\cdot10^{28}\,{\rm G\,cm^2}$), \citet{Caiazzo2021,Ferrario2015} for magnetic white dwarfs, and \citet{GuillochonMcCourt2017} for partial-disruption remnants ($\Bs^{\rm pol}$ enhanced by $\sim 20$ after a prior grazing encounter).
}
\end{table*} 

\subsection{Enhanced stellar magnetic fluxes}

So far we have considered only stellar magnetic fields that can be directly observed.  Here we consider more speculative possibilities, in which {\it unobserved} field could be significantly stronger than what we see.

\subsubsection{Dynamical augmentation of the ``effective'' flux}

During a TDE, roughly half of the stellar mass becomes unbound and escapes to large distances \citep{Rees1988}, while the bound debris falls back toward the black hole.  Field lines may become stretched over very large distances as the bound matter carries them around its orbit while the portion attached to unbound matter moves farther and farther away.   The same effect can stretch field lines even when they penetrate only bound matter when the binding energy of the matter through which they pass varies substantially, so that the times at which some parts of the field lines are brought to the black hole are much shorter than for other parts.
From the perspective of the forming accretion flow, such configurations may behave as effectively open flux systems rather than locally canceling closed loops.  In this fashion, the flux in the vicinity of the black hole can be effectively greater than the true flux.

\citet{Abolmasov2025} showed that the magnetic field intensity can be amplified by 1 -- 2 orders of magnitude over a time $\sim t_0$ as the debris travels towards the BH.  If some of this amplification is reflected in the field intensity on the horizon, $L_{\rm BZ}$ could be a few orders of magnitude larger than the straightforward flux estimate would indicate. This could enable the extreme stars in Table~\ref{tab:stellar-flux-summary} to reach the observed $\LBZ\sim10^{44}\,\ergs$.

\subsubsection{Toroidal Fields}

The stellar magnetic fields we measure are those at the surface.
However, stars may contain a stronger toroidal component below the surface, much harder to observe; an example has been recently discovered
(\citet{Takata2026}, discussed below).
Estimating its strength requires indirect constraints from stability arguments and magnetic evolution models. Despite being observationally elusive, the toroidal component can in principle be much larger than the surface field (up to a factor of $\sim 10^3$ \citep{Braithwaite2009}), and may therefore represent a significant reservoir of magnetic field.

In an intact star, the toroidal field is arranged in closed loops that do not contribute net flux through a spherical surface, and thus does not directly contribute to $\Phis$. 
However, the field line ``stretching'' phenomenon described in the previous subsection can also be applied to toroidal fields, particularly because field loops that are toroidal with respect to the star's center may be rotated to become poloidal with respect to the black hole.

We parametrize the potential toroidal-field contribution
in terms of the ratio $\Bs^{\rm tor}/\Bs$, so that
the Blandford--Znajek luminosity increases $\propto (\Bs^{\rm tor}/\Bs)^2$.
Even allowing optimistic amplification up to $\Bs^{\rm tor}/\Bs\sim10^3$ \citep{Braithwaite2009}, main-sequence and magnetic white dwarfs cannot reach the observed $\LBZ\sim10^{44}\,\ergs$.

Ap/Bp and magnetic O/B stars host fossil magnetic fields that can relax to stable mixed poloidal--toroidal configurations. MHD simulations show typical equilibria at $\Bs^{\rm tor}/\Bs \sim 2$--$5$ \citep{Braithwaite2004,Braithwaite2006}, yielding jet powers $\LBZ \sim 2\cdot10^{43}$--$2\cdot10^{44}\,\ergs$, marginally reaching the observed range in the most favorable cases.
However, stability analysis permits much more extreme configurations: \citet{Braithwaite2009} shows that strongly toroidally dominated fields with $\Bs^{\rm tor}/\Bs \lesssim 300$ are in principle stable. These could produce $\LBZ \sim 10^{46}$--$10^{47}\,\ergs$, though such values represent theoretical upper limits rather than typical equilibria.

Assuming a standard IMF, A/B stars with $\Ms \gtrsim 1.5$--$2\,\Msun$ constitute only a few percent of newly formed main-sequence stars \citep{Kroupa2001}. Since Ap/Bp stars make up only $\sim 5$--$10\%$ of A/B stars \citep{Lignieres2014,Sikora2019}, the implied fraction of Ap/Bp stars among all main-sequence stars is only $\sim \text{few} \times 10^{-3}$. Thus, even before requiring an internal field that is unusually toroidally dominated, Ap/Bp stars are rare among all main-sequence stars.

A notable example of a strong internal stellar magnetic field is provided by
\citet{Takata2026}, who infer lower bounds on the radial and toroidal field
components in the interior of an F-type star. They find
$B_\star \sim 3.5\,{\rm kG}$ and
$B_\star^{\rm tor}/B_\star \sim 26$, consistent with our estimates.
Combined with the inferred radius, $\Rs \sim 2\Rsun$, the toroidal
component corresponds to a magnetic flux sufficient only to power
a jet with $\LBZ \sim 10^{42}\,\ergs$.

\subsubsection{Stellar Mergers as Sources of Enhanced Magnetic Flux}
\citet{Mandel2015} suggested that jetted TDEs could be powered by the magnetic flux from stellar merger remnants in galactic nuclei. Such merger products may possess magnetic fields substantially stronger than those of ordinary stars.
3D MHD simulations of massive main-sequence stellar mergers show that Kelvin--Helmholtz instabilities and shear flows during coalescence can efficiently amplify magnetic fields and generate merger remnants with magnetic fluxes reaching $\Phi \sim 4\cdot10^{28}\,{\rm G\,cm^2}$ \citep{Schneider2019}. 
Observational indications for strongly magnetized merger remnants have also been reported \citep{Shenar2023}.

With these values for a stellar magnetic field, we obtain a Blandford--Znajek luminosity of $\LBZ \sim 10^{43}\,\ergs$.
Although these values marginally reach the observed luminosities of jetted TDEs, they are fully reached in only the extreme scenarios. It is therefore possible that a low-luminosity minority of jetted TDEs could be supported by binary remnant magnetic flux. 

Population estimates suggest that only $8^{+9}_{-4}\%$ of early-type stars are stellar merger products \citep{deMink2014}. Since early-type stars themselves constitute only a few percent of newly formed main-sequence stars for a standard IMF \citep{Kroupa2001}, this corresponds to an IMF-based estimate of only $\sim 10^{-3}$--$10^{-2}$ massive merger remnants per main-sequence star. This channel is further restricted dynamically: \citet{BradnickMandelLevin2017} found that although binaries driven toward the Hills separation radius frequently merge, only some of the resulting merger products are subsequently tidally disrupted by the massive black hole. Thus, although stellar mergers may provide the required magnetic flux in selected systems, they are unlikely to represent a generic channel for most TDEs.

Even more extreme magnetic field values are seen in binary white-dwarf mergers, where localized internal fields can reach $B \sim 10^{10}$--$10^{11}\,{\rm G}$ \citep{Zhu2015}, or $\Phi \sim 4\cdot10^{28}-4\cdot10^{29}\,{\rm G\,cm^2}$ with $R \sim 10^{9}\,{\rm cm}$. However, WD disruptions are possible only around $\MBH\lesssim \text{a few} \times 10^5\Msun$ black holes, which would further increase $\LBZ\propto\MBH^{-2}$; but these systems are less central to our discussion\footnote{Note that \citet{KrolikPiran2012} suggested WD disruption to explain the fast time scales observed in Swift 1644.}.

\subsubsection{Partial disruption}
Simulations of partial disruptions show that the magnetic flux can be enhanced by a factor of $\sim 20$ during the event \citep{GuillochonMcCourt2017}. Thus, in a second encounter with the SMBH, the remnant could exhibit a field that is up to $20$ times larger. The extra factor of $\sim 400$ in the jet luminosity is still insufficient to power jetted TDEs with most types of stars mentioned above. Outliers, however, can marginally reach observed luminosities: partially disrupted O/B stars might create jets with $\LBZ \sim 2\cdot10^{45}\,\ergs$, and partially disrupted Ap/Bp stars might produce jets with $\LBZ \sim 4\cdot10^{44}\,\ergs$. 
Nonetheless, such scenarios are not observationally established and are expected to be very rare, as they require finely tuned grazing encounters in which the star repeatedly loses mass without being fully disrupted. Additionally, for our purpose, they require a highly magnetized progenitor as well. Moreover, even if the first disruption enhances the magnetic field, there is no guarantee that this enhanced field survives until the second disruption, which occurs only after a significant delay.

\subsection{Stellar flux summary}

In this section we have assumed that the stellar magnetic field can be represented by a characteristic large-scale field strength $\Bs$ over an area $\sim \pi \Rs^2$, and that it is efficiently advected to the black hole rather than reconnecting, canceling locally, or remaining attached to debris that does not accrete (the last assumption, as discussed in Sec.~\ref{sec:TDEs},  
{may be valid only in rare cases}). For the enhanced-flux scenarios, we have further assumed that field-line stretching can make closed stellar fields behave as effectively open flux, that dynamical amplification is reflected in the horizon-threading flux, and that internal toroidal fields can be converted into poloidal flux with respect to the black hole. All these assumptions were deliberately chosen to be favorable to the stellar-flux channel.

Under these assumptions, extreme stars, with unusually large radii and magnetic field strengths, can reach the lower end of the observed range, $\LBZ\sim10^{44}\,\ergs$, when taking into account the most optimistic enhancements. However, the combined rarity of the required progenitors and the required amplification mechanism makes it unlikely that this is the main channel for the formation of the observed TDE jets. One possible caveat is that the environment of galactic nuclei may preferentially host or produce highly magnetized stars. 

\section{Magnetic Flux from large radii - ``the Lasso Mechanism''}
\label{sec:Lasso}
\label{sec:power-law-regimes}
\citet{TchekhovskoyMetzger2014} suggested that the TDE debris drags an existing magnetic field surrounding the SMBH from the apoapsis of the most bound orbit all the way to the black hole's horizon, where it deposits enough flux to power jets. In other words, its flux-collecting region stretches out to a radius 
$\sim \hat{a}$.
Note that $\hat{a}$, like all lowercase radii, is measured in units of $\rg$.
\citet{KelleyTchekhovskoy2014} showed that the stream can indeed carry the magnetic flux it encounters  \citep[see also ][]{Piran2015SadTch}.
But the question remains:  is there sufficient flux in the regions through which it passes to support a jet of the required luminosity?

\subsection{Astrophysical setup}
We adopt the large-scale magnetic-field model introduced in
Sec.~\ref{sec:enh}, $B(r)=B_0r^{-q}$, and initially assume that this
single power law extends to the horizon. We discuss 
the effect
of a broken power-law at small radii in Sec.~\ref{sec:broken}. 
We don't explore here the question of how such a magnetic field forms and whether it is likely. In particular, we
do not specify the currents
supporting this field, except to assume that they do not by themselves
produce a luminous accretion flow.
We assume that the debris stream is sufficiently ionized throughout its orbit that it is effectively coupled to the magnetic field.

We consider a full disruption with a penetration factor $\hat{\beta} = 1$, so that $r_p = r_t$, where $r_t$ is given by Eq.~\eqref{eq:r_t_correction}. The debris orbit is approximated as a Keplerian ellipse with pericenter $r_p$, apoapsis $\hat{a}$,
semi-major axis $a_0\approx \hat{a}/2$,
semi-minor axis $b$, and eccentricity
$\varepsilon$. All these lengths are normalized to $\rg$. 

It is important to note that in the following, we approximate the flux as that carried by the fastest returning stream. Although matter from higher energy trajectories will bring in more flux at later times, it will not reach the horizon of the SMBH in time to power the immediate jets we see in TDEs, and thus is not taken into account. Matter arriving from 
$\sim 3 \hat{a}$ 
reaches the black hole $\sim 2 t_0 $ after matter arriving at $t_0$, {a time-dependence that does not match the observed light curves of TDE jets} 
\citep{Gezari2021}. 
Even if matter from higher energy trajectories could bring in flux, it would enhance the amount by only a factor $\lesssim 3^{2-q}$.
Thus, we consider magnetic fields only within $r \leq \hat{a}$.

\subsection{Magnetic flux collected by the debris}
\label{sec:Lasso_flux}

To use the formalism of Sec.~\ref{sec:enh},
we define the characteristic radius as
the semilatus rectum of the debris' elliptical orbit
\begin{equation}
r_c = \frac{b^2}{a_0} = (1+\varepsilon)r_p 
\end{equation}
because the distance from the focus of the ellipse to points on its perimeter is
\begin{equation}
R(\theta) = \frac{r_c}{1 + \varepsilon \cos\theta}
\end{equation}
for azimuthal angle $\theta$.
The covering factor is then
\begin{eqnarray}
\label{eq:fcov_Lasso_def}
 f_{\rm cov}
&=&\frac{r_c^{q-2}}{\pi}
\int_0^{2\pi} \int_{R(\theta) / \rg - \sigma(\theta,\varepsilon) / 2\rg}^{R(\theta) / \rg + \sigma(\theta,\varepsilon) / 2\rg} r^{-q+1}drd\theta,
\end{eqnarray}
where we define the full-width of the swept region as $\sigma(\theta,\varepsilon)$.  To estimate this width, we make the {\it ansatz} $\sigma(\theta,\varepsilon) = \sigma_0 R(\theta)/R_p$ and choose its characteristic scale $\sigma_0 \sim 5 \Rs$ after measuring the stream's width near $R_p$ in data from the simulations of \citet{Ryu+2023}.
Full GRMHD simulations reveal substantially more complex stream evolution \citep{Ryu+2023,SteinbergStone2024,Price+2024}, making both the {\it ansatz} and the measurement uncertain. Nevertheless, they provide a reasonable basis for order of magnitude estimates, and are possibly overestimates of the swept width.

\begin{figure}
    \centering    \includegraphics[width=\linewidth]{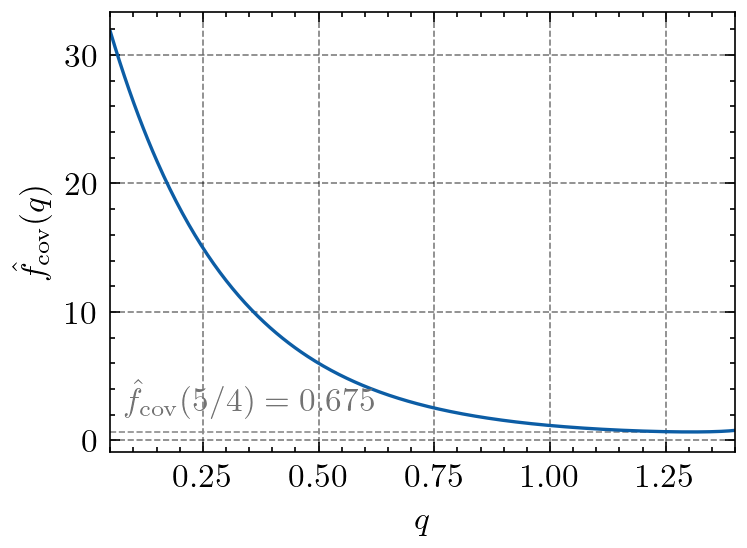}
    \caption{
    $\hat{f}_{\rm cov}$, given in Eq.~\eqref{eq:fhat-closed-form}, for the range $q\in[0.1,1.4]$. The fiducial value is $\hat{f}_{\rm cov}(5/4)\simeq 0.675$.}
    \label{fig:fhat}
\end{figure}

Using the additional assumptions that
$\sigma(\theta,\varepsilon)\ll R(\theta)$ and $\varepsilon\sim1$, we can linearize $f_{\rm cov}$ in terms of $\sigma$, obtaining
\begin{eqnarray}
\label{eq:fcov_Lasso}
f_{\rm cov} &\simeq& \frac{\sigma_0}{\pi \rp} \int_0^{2\pi} \left(1+\varepsilon\cos\theta\right)^{q-2} \,d\theta 
\\ 
\nonumber 
&\equiv&
\hat{f}_{\rm cov}(q)
\left(\frac{\sigma_0}{5\Rs}\right)
\left(\frac{\Xi}{1.47}\right)^{q-\frac{3}{2}}
\left(\frac{\Psi}{0.5}\right)^{q-\frac{5}{2}}
\\
\nonumber
&\times&
\left(\frac{M_{\rm BH}}{10^6\,M_\odot}\right)^{\frac{1-2q}{6}} \left(\frac{M_\star}{M_\odot}\right)^{\frac{2q-1}{6}}.
\end{eqnarray}
Here $\hat f_{\rm cov}$ represents the purely geometrical part, while the scaling factors portray the dependence on dynamical quantities.
For $q\leq1.4$, 
\begin{eqnarray}
\label{eq:fhat-closed-form}
\hat f_{\rm cov}(q)
\approx 0.1\sqrt{\frac{2}{\pi}}\,
\frac{\Gamma\left(\frac{3}{2}-q\right)}{\Gamma(2-q)}
(0.015)^{q-\frac{3}{2}}.
\end{eqnarray}
$\hat  f_{\rm cov} (q)$ is shown in Fig.~\ref{fig:fhat}. For the fiducial $q=5/4$ (see Sec.~\ref{sec:alphaacc} for why this value is chosen), we obtain 
$\hat f_{\rm cov}(5/4) \simeq 0.68$. 
Note that the integral is dominated by the apocenter, which is why
$\hat f_{\rm cov}$ is considerably larger than $\sigma_0/\rp$.

The enhancement factor depends on only the TDE parameters and $q$: 
\begin{eqnarray}
\label{eq:Lasso-luminosity-ratio}
\enh^2
&\sim&
9\,
\cdot \left[46^{5/2-2q}\right]
\left(\frac{\hat{f}_{\rm cov}(q)}{\hat{f}_{\rm cov}(5/4)}\right)^2\rh^{2q-2}
\\
\nonumber
&\times&
\left(\frac{\sigma_0}{5\Rs}\right)^{2}
\left(\frac{\Rs}{\Rsun}\right)^{4-2q}
\left(\frac{\MBH}{10^6\,\Msun}\right)^{\frac{2q-7}{3}}
\left(\frac{\Ms}{\Msun}\right)^{\frac{4q-5}{3}}
\\
\nonumber
&\times&
\left(\frac{\Psi}{0.5}\right)^{-1}
\left(\frac{\Xi}{1.47}\right)^{2q-3}.
\end{eqnarray}

\begin{figure}
\centering
\includegraphics[width=\linewidth]{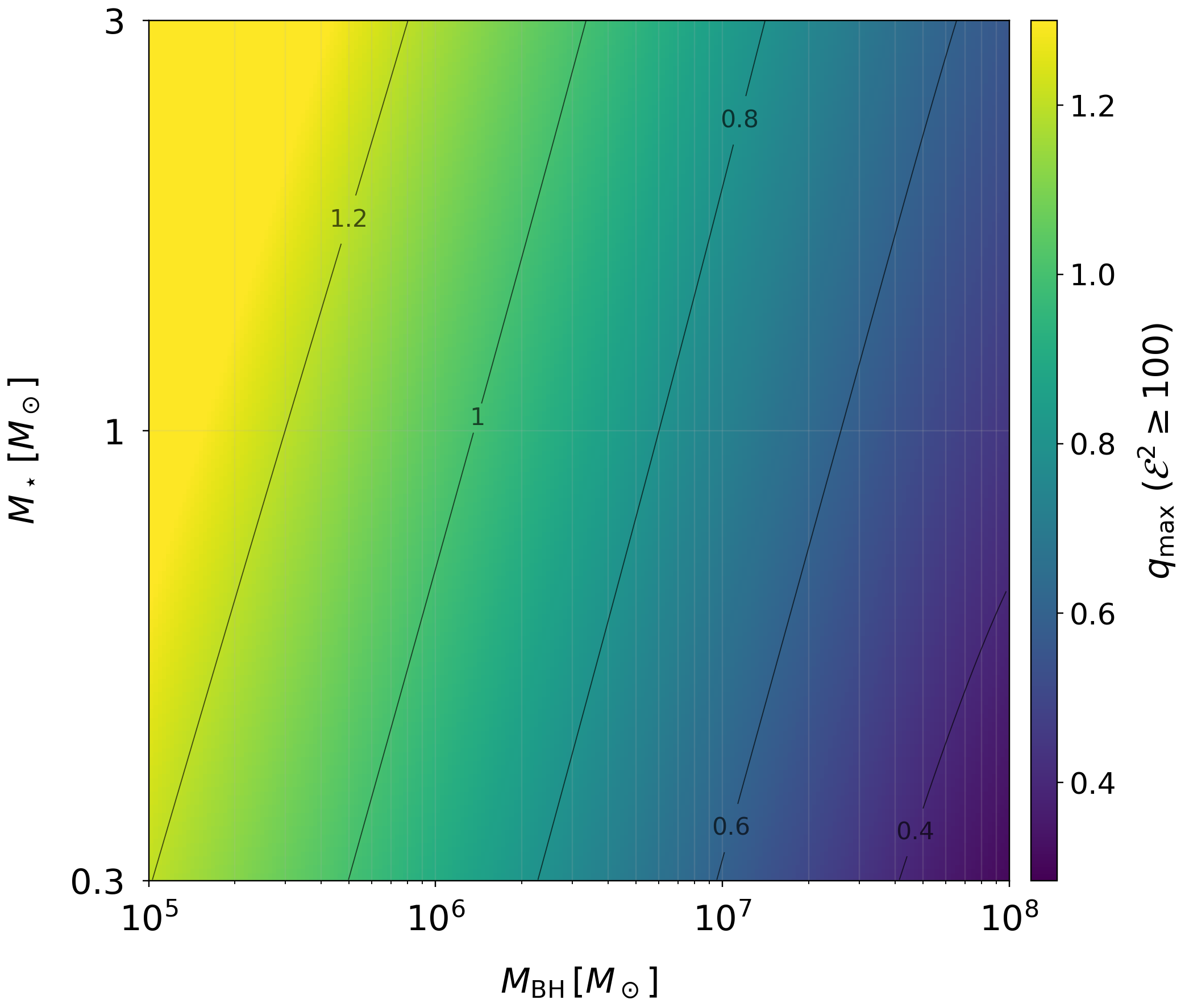}
\caption{Maximum power-law index $q$ for which the Lasso mechanism produces a transient
jet at least two orders of magnitude brighter than the pre-existing steady jet,
$\enh^2\geq100$, as a function of black-hole
and stellar mass. The values are computed from
Eq.~\eqref{eq:Lasso-luminosity-ratio}, using $\rh=1,\sigma_0=5\Rs$ and the stellar
mass-radius relation $\Rs \simeq \Rsun(\Ms/\Msun)^{0.88}$
\citep{Ryu+2020b}.}
\label{fig:q_max_dep}
\end{figure}

Thus, 
for $q=5/4$ and fiducial parameters, the Lasso jet is only 
a factor $\sim 9$ more luminous than 
the pre-existing jet. 
However, the enhancement factor $\enh^2$  
increases rapidly for a flatter magnetic field profile (lower $q$ values).  
For the fiducial TDE parameters, the Lasso model 
can give a jet luminosity at least $100\times$ that of any pre-existing jet
if the magnetic field profile is flatter than $\approx r^{-1}$.
This threshold becomes tighter with increasing black-hole mass or decreasing $\sigma_0$ and loosens with larger stellar mass  (see Fig.~\ref{fig:q_max_dep}).

{We can also estimate the strength of the original field that is great enough to support the observed TDE jet luminosities after the debris collects more flux.}
Using Eqs.~\eqref{eq:Lasso-luminosity-ratio}  and \eqref{eq:LBZ}, we 
find that to reach a given $L_{\rm jet}^{\rm Lasso}$ it must be
\begin{eqnarray}
\label{eq:B0-from-Lasso}
&&B_0 
\sim
1.4\cdot10^5 \,\text{G}\,
\cdot \left[46^{q-5/4}\right]
\left(\frac{L^{\mathrm{Lasso}}_{\mathrm{jet}}}{10^{44} \ergs}\right)^{1/2}
\left(\frac{\sigma_0}{5\Rs}\right)^{-1}
\nonumber
\\
&&\times
\left(\frac{\Rs}{\Rsun}\right)^{q-2}
\left(\frac{M_{\rm BH}}{10^6\,M_\odot}\right)^{(1-2q)/6}
\left(\frac{\Ms}{M_\odot}\right)^{(5-4q)/6}
\\
\nonumber
&&\times
\left(\frac{\hat{f}_{\rm cov}(q)}{\hat{f}_{\rm cov}(5/4)}\right)^{-1}
\left(\frac{\xi_{\rm BZ}}{0.1}\right)^{-1/2}
\left(\frac{\Psi}{0.5}\right)^{1/2}
\left(\frac{\Xi}{1.47}\right)^{\frac{3}{2}-q}.
\end{eqnarray}
This value is 3--4 orders of magnitude  higher than the value inferred from EHT images
of the Galactic center \citep{EHT8}.

Smaller $q$ values increase $\enh^2$
(Eq.~\eqref{eq:Lasso-luminosity-ratio}), but require a more extended magnetic
reservoir around the black hole. We verify in
Appendices~\ref{sec:comparing-thermal-and-magnetic-energy}
and~\ref{sec:pressure-magnetic-pressure} that neither the total magnetic-energy
budget nor the requirement that the debris ram pressure exceed the magnetic
pressure imposes a strong additional constraint on this scenario. 

\subsection{Broken power-law models}
\label{sec:broken}
Both problems that arise in the previous section, high $B_0$ and low $q$, can be alleviated by relaxing the assumption of a single power law.
We consider a broken power-law profile,
\begin{equation}
B(r) = 
\begin{cases}
    B_{0} r_{0}^{-q} (r/r_{0})^{-p}  & r \leq r_{0} \\
    B_{0} r_{0}^{-q} (r/r_{0})^{-q}  & r \geq r_{0},
\end{cases}
\end{equation}
so that the outer region remains unchanged, $B(r)=B_0 r^{-q}$ for $r\ge r_0$, and the profile is continuous at $r_0$. The parameter $p$ controls the inner slope, where 
$p=0$ corresponds to an inner plateau, and $p\gtrless q$ to a steeper or flatter inner slope accordingly.

For $r_{0} \ll \hat{a}$, the transient Lasso jet luminosity $L_{\rm jet}^{\rm Lasso}$ is largely unaffected, as it is dominated by the outer regions. In contrast, the steady-state horizon flux is sensitive to the inner profile: relative to the single power-law case, the field at the horizon scales as $r_0^{\,p-q}$, implying $\enh^2 \propto r_0^{\,2(q-p)}$.
If there is an inner plateau ($p=0$), in which the magnetic field remains approximately constant for $r \lesssim r_0$, the horizon field is suppressed by a factor $\sim r_0^{-q}$, implying $\enh^2\propto r_0^{2q}$. Eq.~\eqref{eq:Lasso-luminosity-ratio} indicates that placing the power-law break at a modest value of $r_0$
would be sufficient for obtaining $\enh^2\sim100$.
However, reducing $B_0$ to match the Sgr A* field intensity would require placing $r_0$ at a much larger multiple of $r_H$, possibly comparable to $\hat a$.

\subsection{Lasso summary}
The Lasso involves gathering magnetic flux from a coherent large-scale ($\sim \hat{a}$) magnetic field. We  considered the case in which its intensity is $\propto r^{-q}$ as an example.
We also assumed that the fastest returning debris carries the relevant flux and that the debris gathers this flux efficiently. We further  assumed that the field is oriented parallel to the debris orbital axis in order to maximize the collected flux, and that the collected flux reaches the black-hole horizon rather than being reconnected on the way.  Most importantly, we assumed that when the debris nears its orbital pericenter, the flux is transferred to the black hole horizon rather than being dragged away by the outward-moving debris, 
an assumption that likely holds only in rare cases, as discussed in Sec.~\ref{sec:TDEs}. 
These assumptions were made to be favorable to the Lasso mechanism; relaxing any of them would reduce the inferred transient jet luminosity, possibly by many orders of magnitude.

With these assumptions, our results demonstrate that for a single power-law distribution
the Lasso model is capable of supplying a sufficiently high transient jet luminosity, but only with a strong  magnetic field ($\gtrsim 1 \times 10^{5} \,\text{G}$) near the horizon.
However, the equally important criterion of {\it not} producing a detectable jet prior to the TDE imposes a strong constraint on the radial-dependence of the magnetic field: the single power-law must be rather flat, $q \lesssim 1$.  The profile can be a bit steeper for $M_{\rm BH} < 10^6 M_\odot$ or $M_* > 1 M_\odot$ (see Fig.~\ref{fig:q_max_dep}).
Satisfying both of these constraints also depends on the quality of our estimate for the characteristic debris stream width, $\sigma_0 \simeq 5 R_*$.  The jet luminosity is $\propto \sigma_0^2$, so the bounds on both the value of $B_0$ required to match observed jet luminosities and  field profile parameter $q$ are tightened if $\sigma_0 < 5 R_*$.

When considering a broken power-law profile, we find that the inner plateau could make this mechanism more viable, relaxing the constraints on both flatness and the strength of the magnetic field.
However, there is little physical motivation for
such a structure: EHT polarimetric modeling of Sgr A* finds a mass-weighted magnetic field strength $B(7.3\rg)=26^{+3}_{-4}\,{\rm G}$ \citep{EHT8}, while a later GRMHD analysis of EHT data finds
$B(4\rg)=67^{+8}_{-9}\,{\rm G}$ \citep{Sen2025}, 
These values suggest an inward increase 
in the field strength rather than a clear plateau.
Moreover, 
in a simulation of large-scale Bondi accretion
\citet{Cho+2023} 
found that the magnetic-field becomes steeper in the inner region, with $r_0 \sim 100$, corresponding to $p=2q$.

\section{Accretion Disk Disturbance}
\label{sec:TDEinAcc}
We turn now to consider pre-existing magnetic fields much nearer to the SMBH,
fields found in accretion disks around SMBHs in galactic nuclei, i.e., active galactic nuclei (AGN). A TDE debris stream that strikes such a 
disk can trigger rapid inflow of the disk material interior to the impact radius, as seen in \citet{Chan2019}. There, the stream--disk interaction generates shocks that dissipate orbital energy and redistribute angular momentum, causing the inner disk to drain rapidly toward the black hole. This inward flow advects magnetic flux associated with the accretion disk toward the horizon. 

Like the Lasso model, this case, too, can be treated with
the formalism developed in  Sec.~\ref{sec:enh}. To analyze this mechanism we leave the index $q$ unspecified at first, later specializing
to a thin-disk $\alpha$-model that fixes $q=5/4$ and $B_0$ (Sec.~\ref{sec:alphaacc}).
Here, the covering factor and characteristic radius 
are $f_{\rm cov} = 1$ and the impact radius.

Defining $\hat{\beta}$ as the penetration factor, i.e. the ratio between $r_t$ and the impact radius, the enhancement factor is given by
\begin{eqnarray}
\label{eq:LjetTDEratio}
\enh^2
&\sim&
49\,
\cdot\left[23^{5/2-2q}\,\frac{9/16}{(2-q)^2}\right]
\rh^{2q-2}
\hat{\beta}^{2q-4} 
\left(\frac{\Rs}{\Rsun}\right)^{4-2q}
\nonumber
\\
&\times&
\left(\frac{\MBH}{10^6\,\Msun}\right)^{\frac{4q-8}{3}}
\left(\frac{\Ms}{\Msun}\right)^{\frac{2q-4}{3}}
\left(\frac{\Psi}{0.5}\right)^{4-2q}.
\end{eqnarray}
Solving Eq.~\eqref{eq:LjetTDEratio} for the value of $\hat{\beta}$
required to obtain $\enh^2\sim100$ gives
\begin{eqnarray}
\label{eq:rminjet}
\hat{\beta}
&\sim&
0.62
\left[
37^{\frac{5/4-q}{2-q}}
\left(
\frac{3/4}{2-q}
\right)^{\frac{1}{2-q}}
\right]
\left(\frac{\enh}{10}\right)^{-\frac{1}{2-q}}
\\
\nonumber
&\times&
\left(\frac{\Rs}{\Rsun}\right)
\left(\frac{\MBH}{10^6\,\Msun}\right)^{-2/3}
\left(\frac{\Ms}{\Msun}\right)^{-1/3}
\left(\frac{\Psi}{0.5}\right)
\rh^{\frac{q-1}{2-q}} .
\end{eqnarray}
For our canonical parameters, an impact radius larger than $\rt$  by about $1.6$ is sufficient for $\enh^2\sim100$.
\citet{Chan2019} find that material originating at larger radii than the impact radius can reach the black-hole horizon, while Tripto et al. (in preparation) demonstrate the associated magnetic-flux transport explicitly. Alternatively, other factors can also operate to increase $\enh$. The disk's orbital timescale at the impact radius is just a few hours.  The stream-disk interaction could rapidly rearrange the pre-existing field into a geometry more favorable for jet production, or,
by stirring turbulence,
may amplify the magnetic field;
amplification by only a factor of $\sim 2 - 3$ would be sufficient for this mechanism to reach ${\cal E}^2 \sim 100$.

Using Eqs.~\eqref{eq:LjetTDEratio} and ~\eqref{eq:LBZ}, we obtain the magnetic field strength $B_0$ required to power a jet with a given luminosity:
\begin{eqnarray}
&&B_0
\sim
6.2\cdot10^4\,{\rm G}\,
\left[23^{q-5/4}\frac{(2-q)}{3/4}\right]
\\
\nonumber
&&\times
\left(\frac{L_{\rm jet}^{\rm dist}}{10^{44}\,\ergs}\right)^{1/2}
\left(\frac{\xi_{\rm BZ}}{0.1}\right)^{-1/2}
\left(\frac{\MBH}{10^6\Msun}\right)^{\frac{1-2q}{3}}
\\
\nonumber
&&\times
\left(\frac{\Rs}{\Rsun}\right)^{q-2}
\left(\frac{\Ms}{\Msun}\right)^{\frac{2-q}{3}}
\left(\frac{\Psi}{0.5}\right)^{q-2}
\hat{\beta}^{2-q} \ , 
\end{eqnarray}
which is comparable to magnetic fields inferred in AGN accretion flows.

The enhancement factor $\cal E$ determines whether the transient can outshine a pre-existing BZ jet. A complementary question is whether the existing disk itself in the pre-TDE state is bright enough to be observed. 
To answer this question, we describe the luminosity in Eddington units.
Imposing a nominal luminosity limit of $L_{\rm acc}^{s} \lesssim L_{\rm limit}^s$ gives $\dot{M} \lesssim 0.1 \dot{M}_{\rm Edd,6} \, (L_{\rm limit}^s / 10^{43} \ergs)$. This limit on the accretion rate also limits the magnetic flux available in the disk. However, as shown in Sec.~\ref{sec:alphaacc}, disks with $\dot{M}\sim10^{-3}$--$10^{-2}\dot{M}_{\rm Edd}$ can contain enough flux to power a strong transient jet.

We now consider two limiting accretion models that span the range of angular-momentum configurations: an $\alpha$-disk and a quasi-spherical Bondi flow. Both are relevant for TDE-triggered jet production: the former represents flows where magnetic amplification can be efficient, while the latter captures low-angular-momentum systems where field reordering is required. 
\\

\subsection{\texorpdfstring{Shakura--Sunayev $\alpha$}{alpha}-disk}
\label{sec:alphaacc}
The $\alpha$-disk model \citep{Shakura1973} provides a simple analytically solvable model for the magnetic profile within a steady-state accretion disk.
We will assume the disk is thin, so that $h \ll R$, where $h$ is the scale height and $R$ is the radius. The stress is given by $t_{\phi r} = -\alpha P_{\rm thermal}$, where $\alpha$ is generally guessed to be $\sim 0.01 - 1$ and independent of radius. 
We will also assume that the magnetic pressure
is a constant fraction of the thermal, i.e., gas plus radiation, pressure, so $B^2/8\pi = \beta^{-1} P_{\rm thermal}$, where $\beta$ is the plasma $\beta$ parameter. 
In traditional 
models, the volume-averaged plasma $\beta$ was typically expected to be $\gg 1$ \citep{BalbusHawley1998}, and global disk simulations including only gas pressure found $\beta \sim O(10)$ in both Newtonian and relativistic treatments \citep{HK2001,DeVilliers+2003}. Recent radiation-MHD simulations, however, often find substantially stronger magnetization in the inner disk, with magnetic pressure approaching the thermal pressure, so that $\beta\sim O(1)$ \citep{BegelmanArmitageReynolds2015,SadowskiNarayan+2014,Jiang2019,Huang+2023b,Zhang+2026}. Therefore we adopt the fiducial value $\beta=2$.

Using the $\alpha$-disk relations, we write the thermal pressure as $P_{\rm th} = \dot{M}\Omega/4\pi h \alpha$.
The magnetic-field profile 
follows immediately: 
\begin{eqnarray}
\label{eq:BRSS} 
B(r)
&=&
\sqrt{8\pi\beta^{-1}P_{\rm th}}
= \left(\frac{2 \dot{M} c}{\bar{h} \alpha \beta \eta\rg^2}\right)^{1/2} r^{-5/4} \ , 
\\
\nonumber
&\sim& 
6.2\cdot10^5\,{\rm G}\,
\left(\frac{\bar{h}}{0.1}\right)^{-1/2}
\left(\frac{\alpha}{0.05}\right)^{-1/2}
\left(\frac{\beta}{2}\right)^{-1/2}
\\
\nonumber
&\times& 
\left(\frac{\eta}{0.1}\right)^{-1}
\left(\frac{\dot{M}}{10^{-3}\dot{M}_{\rm Edd,6}}\right)^{1/2}
\left(\frac{\MBH}{10^6\Msun}\right)^{-1}
r^{-5/4},
\end{eqnarray}
where $\bar{h} \equiv h/R$ and $\eta$ is the rest-mass radiative efficiency. 
This argument gives the fiducial $q=5/4$. 

The central field strength just derived
yields the TDE jet luminosity 
\begin{eqnarray}
\label{eq:Ljet-alpha-TDE}
L_{\rm jet, \alpha}^{\rm dist}
&=&
\frac{64}{9}\,\frac{\xi_{\rm BZ}}{\bar{h}\,\alpha\,\beta}\,
\dot{M} c^2
\left(\frac{r_t}{\hat{\beta}}\right)^{3/2} 
\\
\nonumber
&\sim&
10^{46} \, \ergs
\left(\frac{\xi_{\rm BZ}}{0.1}\right)
\left(\frac{\eta}{0.1}\right)^{-1}
\left(\frac{\bar{h}}{0.1}\right)^{-1}
\left(\frac{\alpha}{0.05}\right)^{-1}
\\
\nonumber
&\times&
\left(\frac{\beta}{2}\right)^{-1}
\left(\frac{\dot{M}}{10^{-3}\,\dot{M}_{\text{Edd},6}}\right)
\left(\frac{\Psi}{0.5}\right)^{3/2}
\hat{\beta}^{-3/2}
\\
\nonumber
&\times&
\left(\frac{\MBH}{10^6\,\Msun}\right)^{-1}
\left(\frac{\Rs}{\Rsun}\right)^{3/2}
\left(\frac{\Ms}{\Msun}\right)^{-1/2}.
\end{eqnarray}
Thus, for our fiducial parameters,
this model predicts 
a jet luminosity within, and potentially above, the range observed in jetted TDEs.
Even if the plasma $\beta$ were as large as found in earlier simulations with cruder treatment of thermodynamics, the jet luminosity would still be $\sim 10^{45}$~erg~s$^{-1}$
for our fiducial parameters, a level barely reached by rare subclasses of stars.
The strongly magnetized inner disks found in recent simulations
provide conditions even more favorable for TDE-triggered magnetic-flux transport and jet production than suggested by older calculations.

Because the characteristic magnetic pressure in this model is assumed to obey $P_B \propto P_{\rm th}$, 
the disturbed-disk jet luminosity scales as $L_{\rm jet}^{\rm dist}\propto B_0^2\propto \dot{M}$.
The steady pre-TDE luminosity is also $\propto \dot M$, so 
the observational limit $L_{\rm acc}^{s}\lesssim L_{\rm limit}^{s}$ also bounds the disturbed-disk jet power. Imposing this constraint gives
\begin{eqnarray}
\label{eq:Ljet-dist-Lacc-limit}
L_{\rm jet}^{\rm dist}
&\lesssim&
7.5\cdot10^{47}\,\ergs\,
\left[
23^{5/2-2q}\frac{9/16}{(2-q)^2}
\right]
\\
\nonumber
&\times&
\left(\frac{\bar{h}}{0.1}\right)^{-1}
\left(\frac{\alpha}{0.05}\right)^{-1}
\left(\frac{\beta}{2}\right)^{-1}
\left(\frac{\eta}{0.1}\right)^{-1}
\left(\frac{\xi_{\rm BZ}}{0.1}\right)
\\
\nonumber
&\times&
\hat{\beta}^{2q-4}
\left(\frac{\Rs}{\Rsun}\right)^{4-2q}
\left(\frac{\MBH}{10^6\,\Msun}\right)^{\frac{4q-8}{3}}
\\
\nonumber
&\times&
\left(\frac{\Ms}{\Msun}\right)^{\frac{2q-4}{3}}
\left(\frac{\Psi}{0.5}\right)^{4-2q}
\left(\frac{L_{\rm limit}^{s}}{10^{43}\,\ergs}\right).
\end{eqnarray}
Thus, the pre-existing accretion luminosity limit still allows high jet luminosities.

Higher jet luminosities are obtainable within this conservative estimate, but we must have a higher accretion rate in the pre-event disk or take into account additional flux advection from outer parts, as discussed above.

\subsection{Bondi accretion}
\label{sec:bondiacc}
\citet{Kwan2023} showed that adding a finite amount of angular momentum to an otherwise Bondi-like flow can reorganize its disordered magnetic field into a coherent structure capable of launching a steady jet. In their simulations, this transition occurs when the gas supplied at the outer boundary has enough specific angular momentum to circularize at $r_{\rm circ}\sim10$--$30$. More generally, simulations find that nearly radial, low-angular-momentum flows produce disordered fields and weak or intermittent jets, whereas modest rotation promotes coherent poloidal flux and stronger jets \citep{Ressler2021, Kwan2023, Lalakos2024, Galishnikova2025}.

It is not clear, however, whether such an idealized Bondi flow can exist within a few tens of $\rg$ at the accretion rates of interest, $\dot M\sim(10^{-3}$--$10^{-2})\dot M_{\rm Edd}(\MBH)$. Gas supplied by stars in a nuclear star cluster is expected to inherit a
characteristic specific angular momentum of order
$
\sim 3\cdot10^{25}\,{\rm cm^2\,s^{-1}},
$
for gas supplied by stars at characteristic radii of $\sim0.5 \,{\rm pc}$, with a stellar velocity dispersion of $\sim200 \,{\rm km~s^{-1}}$ \citep{Cuadra2006,Cuadra2008,Davies2012}.
However, circularization at the relevant impact scale requires only $\sim4\rg c\sim2\cdot10^{22}\,{\rm cm^2\,s^{-1}}$ for $\MBH=10^6\Msun$. The inflow would therefore require very efficient angular-momentum filtering before reaching the black hole. There is a related uncertainty in its magnetization: unlike an $\alpha$-disk, a Bondi flow has no orbital shear to drive the MRI. Radial compression can amplify an existing field, but cannot generate net magnetic flux.

If, nevertheless, magnetic pressure remains a fixed fraction of the ram pressure, then well inside the Bondi radius the free-fall speed $v=c(2/r)^{1/2}$ and steady spherical mass conservation give $\rho=\dot M/(4\pi R^2v)\propto r^{-3/2}$. With $B^2/(8\pi)=\beta^{-1}\rho v^2$, one finds $q=5/4$ and
\begin{eqnarray}
\label{eq:BBondi}
B_0
&=&
\left(\frac{2\sqrt{2}\,\dot M c}{\beta\rg^2}\right)^{1/2}
\sim
5.2\cdot10^4\,{\rm G}\,
\left(\frac{\beta}{2}\right)^{-1/2}
\\
\nonumber
&&\times\left(\frac{\eta}{0.1}\right)^{-1/2}
\left(\frac{\dot{M}}{10^{-3}\dot{M}_{\rm Edd,6}}\right)^{1/2}
\left(\frac{\MBH}{10^6\Msun}\right)^{-1}.
\end{eqnarray}
This is comparable to the value in Eq.~\eqref{eq:B0-from-Lasso}, required to power a $10^{44}\,\ergs$ jet, and it even rises to $\simeq1.7\cdot10^5\,{\rm G}$ at $\dot M=10^{-2}\dot M_{\rm Edd}$.

The returning debris can supply the angular momentum needed to change the flow. Its specific angular momentum, $j_{\rm TDE} \sim \sqrt{2GM_{\rm BH}\rp}$,
corresponds to a circularization radius of
\begin{eqnarray}
r_{\rm circ,TDE}
&\equiv&
\frac{j_{\rm TDE}^2}{GM_{\rm BH}\rg}
\sim
2r_p
=
\frac{2r_t}{\hat{\beta}}
\\
\nonumber
&\sim&
\frac{47}{\hat{\beta}}
\left(\frac{\Rs}{\Rsun}\right)
\left(\frac{\MBH}{10^6\Msun}\right)^{-2/3}
\left(\frac{\Ms}{\Msun}\right)^{-1/3}
\left(\frac{\Psi}{0.5}\right).
\end{eqnarray}
For fiducial parameters this exceeds the threshold found by \citet{Kwan2023} and thus can suffice to produce a strong jet.

The total angular momentum carried by the debris, not only its specific angular momentum, should also be sufficient to dominate that of the pre-existing flow. We show this by comparing the debris mass passing through pericenter during the peak-return interval, $\sim\Ms/3$, with the mass processed by the pre-existing flow over the same time, $\sim\dot M t_0$. Using Eq.~\eqref{eq:t_0_correction},
\begin{eqnarray}
\label{eq:Bondi-mass-test}
\frac{\Ms/3}{\dot M t_0}
&\sim&
2.4\cdot10^5
\left(\frac{\eta}{0.1}\right)
\left(\frac{\dot{M}}{10^{-3}\dot{M}_{\rm Edd,6}}\right)^{-1}
\left(\frac{\MBH}{10^6\Msun}\right)^{-1/2}
\\
\nonumber
&&\times
\left(\frac{\Rs}{\Rsun}\right)^{-3/2}
\left(\frac{\Ms}{\Msun}\right)^2
\left(\frac{\Xi}{1.47}\right)^{3/2}.
\end{eqnarray}
Even at $\dot M=10^{-2}\dot M_{\rm Edd}$, the returning mass exceeds the mass processed by the inner flow by more than four orders of magnitude. Consequently, only a small coupling fraction is needed for the debris to dominate the flow's angular-momentum budget and potentially reorganize its field. Whether this reorganization efficiently produces a coherent jet is left for further numerical work.

\subsection{Accretion summary}

The estimates in this section rely on comparatively few assumptions; the principal one is that the vertical component of the magnetic field accounts for a significant fraction of the total field energy. Hydrodynamic simulations \citep{Chan2019} have shown that the inner disk is swept rapidly to the black hole as a result of TDE debris impact; that it carries its associated magnetic flux along is demonstrated by new MHD simulations (Tripto et~al., in preparation).  That the vertically-integrated magnetic energy density scales roughly with the vertically-integrated pressure was proposed by \citet{Shakura1973}, and, at least in approximate terms, supported by recent simulations combining MHD physics with radiation transfer \citep{Jiang2019,Huang+2023b,Zhang+2026}.  Moreover, none of the fiducial parameters appearing in the estimates of this sections were permitted to vary; their values were all chosen as the usual ones found in the literature.

Under these very limited assumptions, fiducial $\alpha$-disk parameters yield $L_{\rm jet}^{\rm dist}\sim 10^{46}\,\ergs$, indicating that a pre-existing accretion flow can provide the magnetic flux required for a powerful TDE jet. The nominal ratio between the transient and steady jet luminosity is $\sim 50$, but small changes in our model, such as obtaining magnetic flux from radii outside the impact point and positing a more modern accretion disk model would enlarge the possible enhancement factor. 
Thus, it satisfies the criteria we have set with the least reliance on speculation about field enhancement or pre-existing field structure.
In Bondi-like accretion, incoming TDE debris can inject angular momentum into an initially low-angular-momentum flow, potentially reorganizing a disordered magnetic field into a coherent, jet-producing configuration. However, the Bondi accretion model is based on two significant assumptions: first, having purely radial accretion so close to the black hole; and second, having a large magnetic field in that region. Both require further exploration. 

\section{Discussion}
\label{sec:discussion}
\subsection{Lasso vs. accretion models}
\label{sec:Newlassovsaccretion}

As we have already remarked, there are several ways these two mechanisms resemble each other.  Both require substantial magnetic flux to be present around the black hole before the disruption, even though jetted TDEs are not observed to be comparably luminous before the event
\citep{Bloom+2011,Burrows+2011}. 
This fact is what requires our criterion for pre-event and post-event luminosity contrast.  To satisfy this criterion, the hosts must be weakly
accreting or recently active nuclei: faint enough to satisfy the pre-event
limits, yet capable of storing large-scale flux, as in the fossil-disk
picture of \citet{KelleyTchekhovskoy2014}.

There are further formal resemblances.
In the language of Sec.~\ref{sec:enh}, the Lasso (Sec.~\ref{sec:Lasso}) and the
disturbed disk (Sec.~\ref{sec:TDEinAcc}) are two 
ways that
magnetic flux from larger distances can be suddenly brought to the black hole's horizon.
The disturbed disk collects with unit covering
fraction from a modest radius: $f_{\rm cov}=1$ and $r_c\sim{\rm
few}\times10$, so its enhancement factor (Eq.~\ref{eq:LjetTDEratio}) is 
fixed by the impact radius alone, which in turn depends on the original stellar trajectory. 
The Lasso reaches a thousand times
farther, $r_c\sim \hat{a}\sim{\rm few}\times10^3$, but covers only a narrow ribbon
of the annuli it spans, $f_{\rm cov}\sim0.1$, so its enhancement
factor (Eq.~\ref{eq:Lasso-luminosity-ratio}) retains a dependence on the
stellar and black-hole parameters through the stream geometry.
Otherwise the two share the same framework: a pre-existing field $B(r)=B_0 r^{-q}$ extending
to the horizon, coherent advection of the collected flux, and a field
normalization bounded by the requirement that the magnetic energy lie below
the thermal energy of the flow that supports it (expressed through $\beta>1$
in the disk model, and through
Appendix~\ref{sec:comparing-thermal-and-magnetic-energy} for the Lasso).

Their similar mathematical structures lead to cancellations in the ratio of their luminosities,
$L_{\rm jet}^{\rm Lasso}/L_{\rm jet}^{\rm dist}=
(\enh_{\rm Lasso}/\enh_{\rm dist})^2$:
\begin{eqnarray}
&&\left(\frac{\enh_{\rm Lasso}}{\enh_{\rm dist}}\right)^2
\sim
0.18
\left[
2^{5/2-2q}
\frac{(2-q)^2}{9/16}
\right]
\left(\frac{\hat{f}_{\rm cov}(q)}{\hat{f}_{\rm cov}(5/4)}\right)^2
\nonumber
\\
&&\times
\left(\frac{\sigma_0}{5\Rs}\right)^{2}
\left(\frac{\MBH}{10^6\Msun}\right)^{(1-2q)/3}
\left(\frac{\Ms}{\Msun}\right)^{(2q-1)/3}
\\
\nonumber
&&\times
\left(\frac{\Psi}{0.5}\right)^{-1}
\left(\frac{\Xi}{1.47}\right)^{2q-3}
\hat{\beta}^{4-2q}.
\end{eqnarray}
Thus, the Lasso jet luminosity is generically similar to, but somewhat smaller than, the disturbed disk jet luminosity for the same $B_0$.
In terms of
Eq.~\eqref{eq:enh-scaling}, the Lasso's far greater reach,
$(\hat{a}/(r_t/\hat{\beta}))^{2-q}\sim10^{1.5-2}$ for the relevant $q$, is almost
exactly offset by its small covering fraction and by the thinning of the
stream encoded in $\hat{f}_{\rm cov}(q)$.  Larger $M_{\rm BH}$ reduces the ratio $L_{\rm jet}^{Lasso}/L_{\rm jet}^{\rm dist}$; larger $M_*$ increases it.

The near-coincidence of the two enhancement factors should nevertheless be interpreted with caution: the Lasso estimate rests on
some strong and
deliberately favorable assumptions:
full delivery of flux entrained by the debris to the black hole horizon, a shallow radial profile for the pre-existing magnetic field, a relatively high field intensity near the black hole, an uncertain estimate of the stream width, and a field direction that has a significant vertical component.
Sec.~\ref{sec:Lasso}).
In sharp contrast, the only major assumption made in the disturbed disk case is the last one in the preceding list.

Another significant difference difference between the two mechanisms is what is fixed and what is free.
In the disturbed disk, $\enh$  does not depend on $B_0$ 
and depends on the TDE
only through $r_t$, but nothing is adjustable: the accretion model predicts
$q=5/4$ and ties $B_0$ to the accretion rate, so the enhancement is set by
geometry alone, $\enh^2\approx 50$ 
at $\hat{\beta}=1$, reaching
$\enh^2\sim 100$ only for $\hat{\beta}\sim0.6$, Eq.~\eqref{eq:rminjet}. In the Lasso, both $q$ and $B_0$ are free (subject to
the energy constraints above), so $\enh\gtrsim10$ is attainable, but only
for shallow profiles, $q\lesssim1$,
and a fairly strong field near the black hole. These conditions may perhaps describe a fossil,
weakly relaxed configuration but they are not the natural expectation for an
accretion-supported field.  Alternatively, the Lasso can satisfy the contrast
requirement with a broken power law: an inner plateau suppresses
$B_H^{\rm s}$ without affecting the collected flux (Sec.~\ref{sec:broken}).
This route is unavailable to the disturbed disk, whose field profile is
maintained by the accretion flow itself. In both mechanisms $\enh$ is capped
by the MAD limit (Sec.~\ref{sec:MAD}).

An additional observational distinction concerns the prompt TDE
emission. In the Lasso
scenario, the debris dynamics are approximately those of an isolated TDE while
the stream collects ambient magnetic flux.  The prompt thermal flare should
therefore resemble that of a "vanilla" TDE, with magnetic-flux
delivery potentially producing a jet.  In the disturbed-disk
scenario, by contrast, the returning stream collides with a pre-existing
accretion disk.  The resulting interaction should dramatically modify the prompt emission \citep{Chan2019,Chan2020,Chan2021}, in addition to powering any jet through the flux advected by
the disturbed disk.  Thus, signatures of stream--disk interaction in the
prompt flare provide a potential way to distinguish the two mechanisms.

\subsection{MAD constraints on flux accumulation}

\label{sec:MAD}
The mechanisms discussed in this paper rely on transporting magnetic flux toward the black hole. However, magnetic flux cannot be accumulated arbitrarily near the horizon. Once the magnetic pressure becomes comparable to the confining stress of the inflow, it becomes progressively more difficult to add to the near-horizon magnetic flux \citep{BHK2009}; when the accretion flow itself is interrupted, this is deemed 
a magnetically arrested disk (MAD) state
\citep{Tchekhovskoy2011}. 

To illustrate this effect, we assume that the confining stress is dominated by the ram pressure of accretion.
The ratio between the magnetic pressure and the confining stress is then characterized by $\Phi_{\rm BH}^2/(\dot M R_g^2 c)$.  The criterion for the MAD state may then be written as
\begin{eqnarray}
\phi_{\rm BH}
\equiv
\frac{\Phi_{\rm BH}}{\sqrt{\dot{M}_{\rm pre} c \rg^2}}
\lesssim
\phi_{\rm MAD},
\end{eqnarray}
where $\dot M_{\rm pre}$ is the pre-TDE accretion rate.

This argument gives an upper bound on the pre-existing horizon field,
\begin{eqnarray}
\label{eq:BH-MAD-bound}
B_0
&\lesssim&
10^5\,{\rm G}\,
\rh^{-1}
\left(\frac{\phi_{\rm MAD}}{30}\right)
\left(\frac{\eta}{0.1}\right)^{-1/2}
\\
\nonumber
&\times&
\left(\frac{\dot M_{\rm pre}}{10^{-3} \dot M_{\rm Edd}}\right)^{1/2}
\left(\frac{\MBH}{10^6\Msun}\right)^{-1/2},
\end{eqnarray}
using the estimate that $\phi_{\rm MAD} \simeq 30$ \citep{Tchekhovskoy2011}.
Given our previous estimate of the field strength required for the Lasso model, this estimate shows that even in the pre-TDE state, the structure must be near MAD.

During the TDE, the Lasso model may predict an increased accretion rate, and the disturbed disk definitely predicts a sharp rise. Because the MAD flux capacity rises with accretion rate, there may then be room for the additional flux.  However, the transient flux is still bounded by the difference between the post-TDE and pre-TDE fluxes. Assuming $\phi_{\rm post} \dot{M}_{\rm post}^{1/2} \gg \phi_{\rm pre} \dot{M}_{\rm pre}^{1/2}$, we find
\begin{eqnarray}
\label{eq:Phi-trans-MAD-room}
\Phi_{\rm post}
\lesssim
\frac{\phi_{\rm post}}{\phi_{\rm pre}}
\sqrt{c \rg^2 \dot{M}_{\rm {post}}}.
\end{eqnarray}
Here $\Phi_{\rm post}$ is the additional flux supplied by either the Lasso or disk-disturbance mechanism, and $\phi_{\rm pre / post}$ is the dimensionless magnetic flux in the pre/post-TDE states respectively. A mechanism-independent upper bound on the TDE jet luminosity follows:
\begin{eqnarray}
\label{eq:Ltrans-MAD-room}
L_{\rm jet}^{\rm post}
&\lesssim&
4\cdot10^{49}\,\ergs\,
\left(
\frac{\phi_{\rm post}}{\phi_{\rm pre}}
\right)^2
\left(\frac{\xi_{\rm BZ}}{0.1}\right)
\left(\frac{\eta}{0.1}\right)^{-1}
\\
\nonumber
&\times&
\left(\frac{\MBH}{10^6\Msun}\right)
\left(\frac{\dot M_{\rm post}}{\dot M_{\rm Edd}}\right).
\end{eqnarray}
This constraint becomes important only when $\dot{M}_{\rm post}\lesssim 10^{-4}$, so this limit is only interesting when a very small fraction of the derbis stream reaches the black hole.

\section{Summary}
\label{sec:summary}

\begin{table*}[ht] \centering \tabletypesize{\small} \renewcommand{\arraystretch}{1.5} \setlength{\tabcolsep}{2pt} \begin{tabular*}{\textwidth}{ @{\extracolsep{\fill}} c c c c } \hline\hline \tcell{\textbf{Channel}} & \tcell{\textbf{Can supply flux?}} & \tcell{\textbf{Main obstacle}} & \tcell{\textbf{Summary}} \\ \hline \tcell{Stellar flux} & \tcell{Only rare, extreme cases} & \tcell{Flux deficit and rarity} & \tcell{Not generic} \\[3pt]
\tcell{Lasso} & \tcell{Maybe} & \tcell{Strong pre-existing field, shallow field profile, and flux delivery} & \tcell{Viable with parameter tuning} \\ \tcell{Disk disturbance} & \tcell{Yes} & \tcell{Sufficient vertical field} & \tcell{Plausible} \\ \hline\hline \end{tabular*} \caption{\label{tab:flux-channel-summary} Comparison of the three magnetic-flux channels considered in this work.} \end{table*}

We examined whether jetted TDEs can be powered by three magnetic-flux reservoirs: the disrupted star, large-scale flux transported inward by the debris stream (the Lasso mechanism), and flux stored in a pre-existing accretion flow. Table~\ref{tab:flux-channel-summary} summarizes our comparative assessment of the three channels.  In increasing order of promise, they are stellar flux, the Lasso mechanism, and our newly-introduced disk disturbance mechanism.

Our primary gauges for the potential of these mechanisms were the luminosity of the jet produced (at least $\sim 10^{44} - 10^{45}\,\ergs$) and a sufficiently large ratio between the TDE jet luminosity and any pre-existing jet ($\gtrsim 100$). The latter constrains both of the last two models.
Thus, any viable channel must deliver a sufficiently large and coherent magnetic flux to the black-hole horizon on the event timescale, while also producing a transient state much brighter than the pre-TDE system.

Among different stellar types, discussed in Sec.~\ref{sec:stellarflux}, even highly magnetized stars fall short of the required fluxes. 
Some highly magnetized stars can become viable sources if they have hidden toroidal fields that are significantly larger than the observed dipolar components or if the field is somehow magnified,  e.g., by repeated partial disruption or stellar mergers. However, the rarity of suitable progenitors makes this channel unlikely. 

Large-radius transport through the Lasso mechanism discussed in Section~\ref{sec:Lasso}, requires a number of special circumstances. Importantly, recent simulations have shown that only a small fraction of the returning TDE debris approaches the black hole horizon promptly, but quick accretion onto the black hole is a prerequisite for the magnetic flux swept up by the debris to be deposited on the horizon quickly. 
In addition, it requires a rather flat (small power-law index) magnetic field radial profile and a pre-existing field intensity near the black hole considerably greater than inferred for other quiescent supermassive black holes (e.g., through EHT observations of Sgr A*).
Broken power-law profiles can 
loosen the first and, to a lesser extent, the second constraint, but at the cost of introducing additional free parameters.
We therefore view the Lasso channel as viable, but only if the large-scale field configuration is as demonstrated and if some way is found for the collected flux to be conveyed quickly to the black hole.

Disturbance of a pre-existing accretion flow in the inner region, discussed in Section~\ref{sec:TDEinAcc}, can naturally induce rapid inward advection and temporary flux accumulation near the black-hole horizon. As shown in Eq.~\eqref{eq:LjetTDEratio}, even without flux advection from outside the impact radius, the luminosity of observed jets is easily attained.  In addition,
the flux brought by the infalling matter is 
very close to the level necessary 
to explain the observed contrast between the transient and steady states. 
The simulations of \citet{Chan2019} and Tripto et~al. (in prep.) show that material from outside the impact point can also join the inflow, increasing the flux delivered to the black hole, 
raising the ratio of TDE-jet to pre-TDE jet luminosity. The only significant concern regarding this scenario is whether the magnetic field can have the right geometry needed to produce the jet.
The prompt emission offers an  observational
discriminant between the Lasso and accretion models.  The Lasso should produce a more-or-less ordinary TDE flare, while the disk disturbance should show sharp differences from ordinary TDE radiation, as discussed above.

Another accretion scenario is spherical accretion, where a low angular-momentum setup results in an unordered magnetic field configuration. Stream-induced disturbance will not cause advection of flux inwards, but can reorganize the magnetic field structure to form a coherent jet.

Although these arguments are insufficiently rigorous to support a clear identification of the mechanism underlying TDE jets, they do present significant guidance as to the relative promise of these three classes of models as well as motivation and guidance how to explore them further.

Magnetic flux supplied directly by the disrupted star is generally insufficient: ordinary stellar progenitors fall short by orders of magnitude, while viable stellar-progenitor scenarios are limited to rare, highly magnetized objects such as Ap/Bp stars or stellar-merger remnants. 
Magnetic flux from large radii can be brought to the vicinity of the black hole by the Lasso mechanism and produce jets, but only {in special conditions and with a mechanism for taking the magnetic flux ``the last mile" to the black hole horizon.
Disruption-induced disturbance of a pre-existing accretion flow seems to be the most promising.  It requires relatively little fine-tuning in order to match the observed jet luminosities and contrast between the conditions prior to the TDE and during the brief life of the jet that is launched.

\section*{Acknowledgment} 
This work was supported by the Simons Foundation SCEECS collaboration (MP-SCMPS-00001470) (NT,TP), by an ERC advanced grant MultiJets (NT,TP), and by a NASA TCAN grant 80NSSC24K0100 (JHK). 

\newpage

\appendix
\section{MAGNETIC FIELDS OF DIFFERENT STELLAR TYPES}
\label{sec:different-stars-toroidal-field}

This section summarizes the stellar type-dependent estimates underlying Table~\ref{tab:stellar-flux-summary}, including representative poloidal field strengths and radii together with toroidal enhancement of the effective field. 

\paragraph{Solar-like main-sequence stars}
Solar-type main-sequence stars host large-scale dynamo-generated fields with typical surface poloidal components $\Bs^{\rm pol} \sim 1\,{\rm G}$ \citep{Stenflo2013}. For a fiducial solar radius ($\Rs \sim \Rsun$), Eq.~\eqref{eq:LBZ_stellar} gives $\LBZ \sim 10^{30}\,\ergs$.
In $\alpha$--$\Omega$ dynamos, differential rotation converts poloidal field into toroidal field through shear amplification \citep{Parker1955, Charbonneau2010}. As a result, global dynamo simulations of solar-type stars commonly find that the internal toroidal field exceeds the large-scale poloidal component by one to two orders of magnitude, even though the surface toroidal field remains comparable to the poloidal field \citep{Browning2006, Petit2008}. Adopting $\Bs^{\rm tor}/\Bs^{\rm pol}\sim 10 - 100$ increases Eq.~\eqref{eq:LBZ_stellar} to
$
\LBZ \sim 4\cdot10^{31} - 4\cdot10^{33}\,{\rm erg\,s^{-1}},
$
which is still far below the observed $\LBZ \sim 10^{44}\ergs$.

\paragraph{Pre--main-sequence (T-Tauri) stars}
Pre--main-sequence T-Tauri stars possess strong, ordered magnetospheres with $\Bs^{\rm pol} \sim 1$--$3\,{\rm kG}$ and inflated radii $\Rs \sim 2$--$3\,\Rsun$ \citep{JohnsKrull2007}. For a representative system ($\Bs^{\rm pol}\sim 3\,{\rm kG}$, $\Rs \sim 3\,\Rsun$), Eq.~\eqref{eq:LBZ_stellar} gives $\LBZ \sim 10^{39}\,\ergs$.
T-Tauri stars host dynamo-generated large-scale magnetospheres with mixed topologies. Spectropolarimetric reconstructions typically find toroidal components comparable to, or moderately exceeding, the poloidal field \citep{Donati2009, Donati2012}. A representative range is $\Bs^{\rm tor}/\Bs^{\rm pol} \sim 0.3 - 10$. This enhances the luminosity estimate to $\LBZ \sim 3 \cdot 10^{38} - 3\cdot10^{40}\,\ergs$.

\paragraph{Ap/Bp stars}
Chemically peculiar Ap/Bp stars host stable, large-scale fossil magnetic fields with typical $\Bs^{\rm pol} \sim 1$--$10\,{\rm kG}$ and $\Rs \sim 2$--$5\,\Rsun$ \citep{Romanyuk2025}. At the extreme end, exemplified by Babcock's star ($\Bs^{\rm pol}\sim 3\cdot10^4\,{\rm G}$, $\Rs\sim 5\,\Rsun$) \citep{Babcock1960_2}, Eq.~\eqref{eq:LBZ_stellar} gives $\LBZ \sim 10^{42}\,\ergs$ for the observed poloidal component alone.
Like other fossil-field stars, Ap/Bp objects can relax to stable mixed poloidal--toroidal equilibria. MHD simulations find typical ratios $\Bs^{\rm tor}/\Bs^{\rm pol} \sim 2$--$5$ \citep{Braithwaite2004,Braithwaite2006}. Adopting an effective toroidal field with this ratio for Babcock's star ($\Bs^{\rm tor}\sim 10^5\,{\rm G}$) yields $\LBZ \sim 10^{43}\,\ergs$, marginally approaching but still below the observed $\LBZ \sim 10^{44}\,\ergs$.

\paragraph{Magnetic O/B stars}
A small fraction of massive O/B-type stars possess strong, globally ordered fossil fields. NGC~1624-2, among the most extreme known examples, has $\Bs^{\rm pol}\sim 2\cdot10^4\,{\rm G}$ and $\Rs\sim 10\,\Rsun$ \citep{David-Uraz2021,Wade2012_2}. For this system, Eq.~\eqref{eq:LBZ_stellar} gives $\LBZ \sim 6\cdot10^{42}\,\ergs$ using the poloidal field alone---the largest ordinary main-sequence flux estimate in Table~\ref{tab:stellar-flux-summary}.
Fossil-field stability arguments and MHD equilibria analogous to those for Ap/Bp stars suggest comparable toroidal enhancement, $\Bs^{\rm tor}/\Bs^{\rm pol} \sim 2$--$5$ \citep{Braithwaite2004,Braithwaite2006}, which would raise $\LBZ$ to $\sim 10^{43}$--$10^{44}\,\ergs$ in the most favorable cases.

\paragraph{Magnetic white dwarfs}
Some magnetic white dwarfs exhibit extremely strong surface fields, $\Bs^{\rm pol} \sim 10^7$--$10^9\,{\rm G}$, but their compact radii ($\Rs \sim 10^8\,{\rm cm}$) limit the total magnetic flux \citep{Caiazzo2021,Ferrario2015}. For a representative strongly magnetized system ($\Bs^{\rm pol}\sim 10^9\,{\rm G}$, $\Rs\sim 10^8\,{\rm cm}$), Eq.~\eqref{eq:LBZ_stellar} gives $\LBZ \sim 7\cdot10^{36}\,\ergs$.
Like other radiative  
stars,
stability requires mixed
poloidal--toroidal configurations.
Hall evolution simulations of white dwarf magnetic fields
typically produce configurations in which the toroidal and
poloidal components are comparable,
corresponding to $\Bs^{\rm tor}/\Bs^{\rm pol} \sim 0.3$--$3$
\citep{Gourgouliatos2014}.
More strongly toroidally dominated equilibria are not
formally excluded by stability arguments, but are not
commonly produced in current simulations. Adopting a representative range $\Bs^{\rm tor}/\Bs^{\rm pol} \sim 0.3$--$3$
gives $\LBZ \sim 10^{39}\,\ergs$ for the most
strongly magnetized systems.
Even if we allow for a moderate toroidal ratio with $\Bs^{\rm tor}/\Bs^{\rm pol} \sim 100$, the maximum stellar magnetic flux of white dwarfs yields $\LBZ \sim 10^{42}\,\ergs$. This is still well below the $\LBZ \sim 10^{44}\,\ergs$ inferred for the most luminous jetted TDEs, and in most cases, far below this level.

\section{Lasso-Stream}
\label{app:Lasso-stream}

\subsection{Comparing Thermal and Magnetic Energy}
\label{sec:comparing-thermal-and-magnetic-energy}
In the original Lasso scenario \citep{TchekhovskoyMetzger2014}, the field is assumed to be a fossil remnant of a previously active nucleus. In this framework, the instantaneous thermal energy stored in the quiescent accretion disk
suggests a natural scale
for the available magnetic energy. 

We first estimate the magnetic energy associated with the required field profile. Since the integrated magnetic energy scales as $U_B(r)\propto r^{3-2q}$, for $q>3/2$ the magnetic energy is
dominated 
by the smallest radii,
while $q=3/2$ gives only a logarithmic dependence on the outer radius.  In addition, in the regime $q\geq 3/2$, $\enh^2 \lesssim 0.1$, which does not match observations.
For all these reasons, we rule out $q \geq 3/2$.
For $q<3/2$, we adopt $\hat{a}$ as the outer radius, which is a conservative estimate because the stream must extend at least up to this scale, and any larger radius would increase the magnetic energy. The magnetic energy,
$ U_B = \int_2^{\hat{a}} [B^2(r)/8\pi] dV$, using Eq.~\eqref{eq:B0-from-Lasso},
is then
\begin{eqnarray}
\label{eq:magenergy}
U_B
&\approx&
3\cdot10^{44}\,{\rm erg}\,
\left[
\frac{100^{5/2-2q}}
{2(3-2q)}
\right]
\left(\frac{\hat{f}_{\rm cov}(q)}{\hat{f}_{\rm cov}(5/4)}\right)^{-2}
\\
\nonumber
&\times&
\left(\frac{\sigma_0}{5\Rs}\right)^{-2}
\left(
\frac{L_{\rm jet}^{\rm Lasso}}
{10^{44}\,{\rm erg}\,{\rm s}^{-1}}
\right)
\left(\frac{\xi_{\rm BZ}}{0.1}\right)^{-1}
\left(\frac{\bar{h}}{0.1}\right)
\\
\nonumber
&\times&
\left(\frac{M_{\rm BH}}{10^6\,M_\odot}\right)^{7/3}
\left(\frac{M_\star}{M_\odot}\right)^{-1/3}
\left(\frac{R_\star}{R_\odot}\right)^{-1}
\left(\frac{\Psi}{0.5}\right).
\end{eqnarray}
where $\bar{h} \equiv h/r$ accounts for the aspect ratio of the accretion flow supporting the magnetic field.
We compare this to the disk thermal energy at luminosity $L_{\rm disk}$, measured in units of the Eddington luminosity $L_{\rm Edd}(\MBH)$. The thermal energy is estimated as 
$E_{\rm th} \sim L_{\rm disk} t_{\rm th}$, where
$t_{\rm th} \sim (\alpha \Omega_K)^{-1}$ is the thermal timescale in a thin $\alpha$-disk. Evaluating it at the characteristic outer radius $\hat{a}$ 
 gives
\begin{eqnarray}
\label{eq:E_th}
E_{\rm th}
&\sim&
1.1 \cdot 10^{51}\,{\rm erg}\,
\left(\frac{L_{\rm disk}}{L_{\rm Edd}}\right)
\left(\frac{\alpha}{0.05}\right)^{-1}
\\
\nonumber
&\times&
\left(\frac{M_{\rm BH}}{10^6 M_\odot}\right)^{3/2}
\left(\frac{M_*}{M_\odot}\right)^{-1}
\left(\frac{\Rs}{\Rsun}\right)^{3/2}
\left(\frac{\Xi}{1.47}\right)^{-3/2}.
\end{eqnarray}

For all $L_{\rm disk} \in [10^{-4}, 1] L_{\rm Edd}$, $\MBH \in [10^5, 10^8] \Msun$, and $\Ms \in [0.3, 3] \Msun$, we find $U_B \lesssim E_{\rm th}$, which means that imposing $E_{\rm th} \gtrsim U_B$ does not constrain the Lasso mechanism.

\subsection{Stream pressure vs. magnetic pressure}
\label{sec:pressure-magnetic-pressure}

For the debris stream to advect magnetic flux inward, its ram pressure must exceed the magnetic pressure, both the local magnetic pressure at every point, given by Eq.~\eqref{eq:B0-from-Lasso}, and the accumulated magnetic pressure in the inflowing debris stream, estimated using the required flux in Eq.~\eqref{eq:Phi-from-LBZ} and the stream-width scaling below. 
For this estimate, we choose the peak mass return rate, $\dot{M}_{\rm stream}\sim(\Ms/3)/t_0$, with the fallback time $t_0$ given by Eq.~\eqref{eq:t_0_correction}, a stream velocity $v_{\rm stream}\sim c \, r^{-1/2}$, and full stream width as in Sec.~\ref{sec:Lasso_flux},
$
\sigma(r)\sim\sigma_0 R/\rp,
$
with $\sigma_0\sim5\Rs$, and $\rp\sim\rt$,
where the tidal radius is given by Eq.~\eqref{eq:r_t_correction}. The stream ram pressure is thus
\begin{eqnarray}
\label{eq:P-ram-local}
&&P_{\rm ram}
\sim
\frac{\dot{M}_{\rm stream}}
{\pi[\sigma(r)/2]^2}
v_{\rm stream}
\\
\nonumber
&&\sim
1.1\cdot10^{16}\,
{\rm erg\,cm^{-3}}\,
r^{-5/2}
\left(\frac{\Psi}{0.5}\right)^2
\left(\frac{\Xi}{1.47}\right)^{-1}
\\
\nonumber
&&\times
\left(\frac{\MBH}{10^6\Msun}\right)^{-11/6}
\left(\frac{\Rs}{\Rsun}\right)^{-3/2}
\left(\frac{\Ms}{\Msun}\right)^{4/3}.
\end{eqnarray}
Comparing Eq.~\eqref{eq:P-ram-local} to the local magnetic pressure
$B_{\rm local}^2(r)/(8\pi)$, with $B_{\rm local}(r)=B_0r^{-q}$ and $B_0$ from
Eq.~\eqref{eq:B0-from-Lasso}, we obtain
\begin{eqnarray}
\label{eq:p_ratio_local}
&&\frac{P_{\rm ram}}{B_{\rm local}^2(r)/(8\pi)}
\sim
1.2\cdot10^9
\left[46^{3/2-2q}\right]
r^{2q-5/2}
\\
\nonumber
&&\times
\left(
\frac{\hat{f}_{\rm cov}(q)}
{\hat{f}_{\rm cov}(5/4)}
\right)^{-2}
\left(\frac{\sigma_0}{5\Rs}\right)^{2}
\left(
\frac{L_{\rm jet}^{\rm Lasso}}
{10^{44}\,{\rm erg\,s^{-1}}}
\right)^{-1}
\\
\nonumber
&&\times
\left(\frac{\MBH}{10^6\Msun}\right)^{(4q-13)/6}
\left(\frac{\Rs}{\Rsun}\right)^{5/2-2q}
\left(\frac{\Ms}{\Msun}\right)^{(4q-1)/3}
\\
\nonumber
&&\times
\left(\frac{\xi_{\rm BZ}}{0.1}\right)
\left(\frac{\Psi}{0.5}\right)
\left(\frac{\Xi}{1.47}\right)^{2q-4}.
\end{eqnarray}

For almost all parameter space, the ram pressure is sufficient to overcome the local magnetic pressure in the region $r\lesssim \hat{a}$. This condition 
is violated
only for very small $q$ values and very large black-hole masses,
$\MBH\gtrsim{\rm few}\cdot10^7\Msun$ and $q\lesssim0.5$.

The ram pressure should also exceed the magnetic pressure of the accumulated field carried by the stream. Over one orbital period, the debris stream collects magnetic flux $\Phi$, given by Eq.~\eqref{eq:Phi-from-LBZ}, distributed uniformly over its cross-sectional area, such that
$B_{\rm acc}(r)
\sim
\Phi/\pi[\sigma(r)/2]^2$.
For $\sigma\propto r$, this implies $B_{\rm acc}\propto r^{-2}$ and therefore
$B_{\rm acc}^2/(8\pi)\propto r^{-4}$. We thus obtain
\vskip -0.5cm
\begin{eqnarray}
\frac{B_{\rm acc}^2(r)}{8\pi}
&\sim&
1.4\cdot10^{17}\,
{\rm erg\,cm^{-3}}\,
r^{-4}
\left(
\frac{L_{\rm BZ}}
{10^{44}\,{\rm erg\,s^{-1}}}
\right)
\\
\nonumber
 &\times&
\left(\frac{\xi_{\rm BZ}}{0.1}\right)^{-1}
\left(\frac{\MBH}{10^6\Msun}\right)^{-2/3}
\left(\frac{\Ms}{\Msun}\right)^{-4/3}
\left(\frac{\Psi}{0.5}\right)^4.
\end{eqnarray}

Comparing to Eq.~\eqref{eq:P-ram-local}, we arrive at
\begin{eqnarray}
\label{eq:p_ratio_accumulated}
&&\frac{P_{\rm ram}}{B_{\rm acc}^2/(8\pi)}
\sim
0.78\,
r^{3/2}
\left(
\frac{L_{\rm BZ}}
{10^{44}\,{\rm erg\,s^{-1}}}
\right)^{-1}
\left(\frac{\xi_{\rm BZ}}{0.1}\right)
\\
\nonumber
&&\times
\left(\frac{\MBH}{10^6\Msun}\right)^{-7/6}
\left(\frac{\Rs}{\Rsun}\right)^{-3/2}
\left(\frac{\Ms}{\Msun}\right)^{8/3}
\\
\nonumber
&&\times
\left(\frac{\Psi}{0.5}\right)^{-2}
\left(\frac{\Xi}{1.47}\right)^{-1}.
\end{eqnarray}

For fiducial parameters, this ratio becomes smaller than unity only very close to the black hole, $r\lesssim1.2$, and thus this does not pose a constraint on the Lasso model.


\begin{thebibliography}{}
\expandafter\ifx\csname natexlab\endcsname\relax\def\natexlab#1{#1}\fi
\providecommand{\url}[1]{\href{#1}{#1}}
\providecommand{\dodoi}[1]{doi:~\href{http://doi.org/#1}{\nolinkurl{#1}}}
\providecommand{\doeprint}[1]{\href{http://ascl.net/#1}{\nolinkurl{http://ascl.net/#1}}}
\providecommand{\doarXiv}[1]{\href{https://arxiv.org/abs/#1}{\nolinkurl{https://arxiv.org/abs/#1}}}

\bibitem[{{Abolmasov} {et~al.}(2025){Abolmasov}, {Bromberg}, {Levinson}, \& {Nakar}}]{Abolmasov2025}
{Abolmasov}, P., {Bromberg}, O., {Levinson}, A., \& {Nakar}, E. 2025, arXiv e-prints, arXiv:2509.23894, \dodoi{10.48550/arXiv.2509.23894}

\bibitem[{{Alexander} {et~al.}(2020){Alexander}, {van Velzen}, {Horesh}, \& {Zauderer}}]{Alexander2020}
{Alexander}, K.~D., {van Velzen}, S., {Horesh}, A., \& {Zauderer}, B.~A. 2020, \ssr, 216, 81, \dodoi{10.1007/s11214-020-00702-w}

\bibitem[{{Alexander} {et~al.}(2017){Alexander}, {Wieringa}, {Berger}, {Saxton}, \& {Komossa}}]{Alexander2017}
{Alexander}, K.~D., {Wieringa}, M.~H., {Berger}, E., {Saxton}, R.~D., \& {Komossa}, S. 2017, \apj, 837, 153, \dodoi{10.3847/1538-4357/aa6192}

\bibitem[{{Andreoni} {et~al.}(2022){Andreoni}, {Coughlin}, {Perley}, {Yao}, {Lu}, {Cenko}, {Kumar}, {Anand}, {Ho}, {Kasliwal}, {de Ugarte Postigo}, {Sagu{\'e}s-Carracedo}, {Schulze}, {Kann}, {Kulkarni}, {Sollerman}, {Tanvir}, {Rest}, {Izzo}, {Somalwar}, {Kaplan}, {Ahumada}, {Anupama}, {Auchettl}, {Barway}, {Bellm}, {Bhalerao}, {Bloom}, {Bremer}, {Bulla}, {Burns}, {Campana}, {Chandra}, {Charalampopoulos}, {Cooke}, {D'Elia}, {Das}, {Dobie}, {Ag{\"u}{\'\i} Fern{\'a}ndez}, {Freeburn}, {Fremling}, {Gezari}, {Goode}, {Graham}, {Hammerstein}, {Karambelkar}, {Kilpatrick}, {Kool}, {Krips}, {Laher}, {Leloudas}, {Levan}, {Lundquist}, {Mahabal}, {Medford}, {Miller}, {M{\"o}ller}, {Mooley}, {Nayana}, {Nir}, {Pang}, {Paraskeva}, {Perley}, {Petitpas}, {Pursiainen}, {Ravi}, {Ridden-Harper}, {Riddle}, {Rigault}, {Rodriguez}, {Rusholme}, {Sharma}, {Smith}, {Stein}, {Th{\"o}ne}, {Tohuvavohu}, {Valdes}, {van Roestel}, {Vergani}, {Wang}, \& {Zhang}}]{Andreoni+2022}
{Andreoni}, I., {Coughlin}, M.~W., {Perley}, D.~A., {et~al.} 2022, \nat, 612, 430, \dodoi{10.1038/s41586-022-05465-8}

\bibitem[{{Babcock}(1960)}]{Babcock1960_2}
{Babcock}, H.~W. 1960, \apj, 132, 521, \dodoi{10.1086/146960}

\bibitem[{{Balbus} \& {Hawley}(1998)}]{BalbusHawley1998}
{Balbus}, S.~A., \& {Hawley}, J.~F. 1998, Reviews of Modern Physics, 70, 1, \dodoi{10.1103/RevModPhys.70.1}

\bibitem[{{Barniol Duran} \& {Piran}(2013)}]{BarniolDuran2013}
{Barniol Duran}, R., \& {Piran}, T. 2013, \apj, 770, 146, \dodoi{10.1088/0004-637X/770/2/146}

\bibitem[{{Beckwith} {et~al.}(2008){Beckwith}, {Hawley}, \& {Krolik}}]{BHK2008}
{Beckwith}, K., {Hawley}, J.~F., \& {Krolik}, J.~H. 2008, \apj, 678, 1180, \dodoi{10.1086/533492}

\bibitem[{{Beckwith} {et~al.}(2009){Beckwith}, {Hawley}, \& {Krolik}}]{BHK2009}
---. 2009, \apj, 707, 428, \dodoi{10.1088/0004-637X/707/1/428}

\bibitem[{{Begelman} {et~al.}(2015){Begelman}, {Armitage}, \& {Reynolds}}]{BegelmanArmitageReynolds2015}
{Begelman}, M.~C., {Armitage}, P.~J., \& {Reynolds}, C.~S. 2015, \apj, 809, 118, \dodoi{10.1088/0004-637X/809/2/118}

\bibitem[{{Berger} {et~al.}(2012){Berger}, {Zauderer}, {Pooley}, {Soderberg}, {Sari}, {Brunthaler}, \& {Bietenholz}}]{Berger2012}
{Berger}, E., {Zauderer}, A., {Pooley}, G.~G., {et~al.} 2012, \apj, 748, 36, \dodoi{10.1088/0004-637X/748/1/36}

\bibitem[{{Blandford} \& {Znajek}(1977)}]{BZ1977}
{Blandford}, R.~D., \& {Znajek}, R.~L. 1977, \mnras, 179, 433, \dodoi{10.1093/mnras/179.3.433}

\bibitem[{{Bloom} {et~al.}(2011){Bloom}, {Giannios}, {Metzger}, {Cenko}, {Perley}, {Butler}, {Tanvir}, {Levan}, {O'Brien}, {Strubbe}, {De Colle}, {Ramirez-Ruiz}, {Lee}, {Nayakshin}, {Quataert}, {King}, {Cucchiara}, {Guillochon}, {Bower}, {Fruchter}, {Morgan}, \& {van der Horst}}]{Bloom+2011}
{Bloom}, J.~S., {Giannios}, D., {Metzger}, B.~D., {et~al.} 2011, Science, 333, 203, \dodoi{10.1126/science.1207150}

\bibitem[{{Bradnick} {et~al.}(2017){Bradnick}, {Mandel}, \& {Levin}}]{BradnickMandelLevin2017}
{Bradnick}, B., {Mandel}, I., \& {Levin}, Y. 2017, \mnras, 469, 2042, \dodoi{10.1093/mnras/stx1007}

\bibitem[{{Braithwaite}(2009)}]{Braithwaite2009}
{Braithwaite}, J. 2009, \mnras, 397, 763, \dodoi{10.1111/j.1365-2966.2008.14034.x}

\bibitem[{{Braithwaite} \& {Nordlund}(2006)}]{Braithwaite2006}
{Braithwaite}, J., \& {Nordlund}, {\r{A}}. 2006, \aap, 450, 1077, \dodoi{10.1051/0004-6361:20041980}

\bibitem[{{Braithwaite} \& {Spruit}(2004)}]{Braithwaite2004}
{Braithwaite}, J., \& {Spruit}, H.~C. 2004, \nat, 431, 819, \dodoi{10.1038/nature02934}

\bibitem[{{Brown} {et~al.}(2015){Brown}, {Levan}, {Stanway}, {Tanvir}, {Cenko}, {Berger}, {Chornock}, \& {Cucchiaria}}]{Brown+2015}
{Brown}, G.~C., {Levan}, A.~J., {Stanway}, E.~R., {et~al.} 2015, \mnras, 452, 4297, \dodoi{10.1093/mnras/stv1520}

\bibitem[{{Browning} \& {Basri}(2006)}]{Browning2006}
{Browning}, M.~K., \& {Basri}, G. 2006, in American Astronomical Society Meeting Abstracts, Vol. 209, American Astronomical Society Meeting Abstracts, 89.08

\bibitem[{{Burrows} {et~al.}(2011){Burrows}, {Kennea}, {Ghisellini}, {Mangano}, {Zhang}, {Page}, {Eracleous}, {Romano}, {Sakamoto}, {Falcone}, {Osborne}, {Campana}, {Beardmore}, {Breeveld}, {Chester}, {Corbet}, {Covino}, {Cummings}, {D'Avanzo}, {D'Elia}, {Esposito}, {Evans}, {Fugazza}, {Gelbord}, {Hiroi}, {Holland}, {Huang}, {Im}, {Israel}, {Jeon}, {Jeon}, {Jun}, {Kawai}, {Kim}, {Krimm}, {Marshall}, {P. M{\'e}sz{\'a}ros}, {Negoro}, {Omodei}, {Park}, {Perkins}, {Sugizaki}, {Sung}, {Tagliaferri}, {Troja}, {Ueda}, {Urata}, {Usui}, {Antonelli}, {Barthelmy}, {Cusumano}, {Giommi}, {Melandri}, {Perri}, {Racusin}, {Sbarufatti}, {Siegel}, \& {Gehrels}}]{Burrows+2011}
{Burrows}, D.~N., {Kennea}, J.~A., {Ghisellini}, G., {et~al.} 2011, \nat, 476, 421, \dodoi{10.1038/nature10374}

\bibitem[{{Caiazzo} {et~al.}(2021){Caiazzo}, {Burdge}, {Fuller}, {Heyl}, {Kulkarni}, {Prince}, {Richer}, {Schwab}, {Andreoni}, {Bellm}, {Drake}, {Duev}, {Graham}, {Helou}, {Mahabal}, {Masci}, {Smith}, \& {Soumagnac}}]{Caiazzo2021}
{Caiazzo}, I., {Burdge}, K.~B., {Fuller}, J., {et~al.} 2021, \nat, 596, E15, \dodoi{10.1038/s41586-021-03799-3}

\bibitem[{{Cendes} {et~al.}(2022){Cendes}, {Berger}, {Alexander}, {Gomez}, {Hajela}, {Chornock}, {Laskar}, {Margutti}, {Metzger}, {Bietenholz}, \& et~al.}]{Cendes+2022}
{Cendes}, Y., {Berger}, E., {Alexander}, K.~D., {et~al.} 2022, \apj, 938, 28, \dodoi{10.3847/1538-4357/ac88d0}

\bibitem[{{Cenko} {et~al.}(2012){Cenko}, {Krimm}, {Horesh}, {Rau}, {Frail}, {Kennea}, {Levan}, {Holland}, {Butler}, {Quimby}, {Bloom}, {Filippenko}, {Gal-Yam}, {Greiner}, {Kulkarni}, {Ofek}, {Olivares E.}, {Schady}, {Silverman}, {Tanvir}, \& {Xu}}]{Cenko+2012}
{Cenko}, S.~B., {Krimm}, H.~A., {Horesh}, A., {et~al.} 2012, \apj, 753, 77, \dodoi{10.1088/0004-637X/753/1/77}

\bibitem[{{Chan} {et~al.}(2020){Chan}, {Piran}, \& {Krolik}}]{Chan2020}
{Chan}, C.-H., {Piran}, T., \& {Krolik}, J.~H. 2020, \apj, 903, 17, \dodoi{10.3847/1538-4357/abb776}

\bibitem[{{Chan} {et~al.}(2021){Chan}, {Piran}, \& {Krolik}}]{Chan2021}
---. 2021, \apj, 914, 107, \dodoi{10.3847/1538-4357/abf0a7}

\bibitem[{{Chan} {et~al.}(2019){Chan}, {Piran}, {Krolik}, \& {Saban}}]{Chan2019}
{Chan}, C.-H., {Piran}, T., {Krolik}, J.~H., \& {Saban}, D. 2019, \apj, 881, 113, \dodoi{10.3847/1538-4357/ab2b40}

\bibitem[{{Charbonneau}(2010)}]{Charbonneau2010}
{Charbonneau}, P. 2010, Living Reviews in Solar Physics, 7, 3, \dodoi{10.12942/lrsp-2010-3}

\bibitem[{{Cho} {et~al.}(2023){Cho}, {Prather}, {Narayan}, {Natarajan}, {Su}, {Ricarte}, \& {Chatterjee}}]{Cho+2023}
{Cho}, H., {Prather}, B.~S., {Narayan}, R., {et~al.} 2023, \apjl, 959, L22, \dodoi{10.3847/2041-8213/ad1048}

\bibitem[{{Cuadra} \& {Nayakshin}(2006)}]{Cuadra2006}
{Cuadra}, J., \& {Nayakshin}, S. 2006, in Journal of Physics Conference Series, Vol.~54, Journal of Physics Conference Series, ed. R.~{Sch{\"o}del}, G.~C. {Bower}, M.~P. {Muno}, S.~{Nayakshin}, \& T.~{Ott} (IOP), 436--442, \dodoi{10.1088/1742-6596/54/1/068}

\bibitem[{{Cuadra} {et~al.}(2008){Cuadra}, {Nayakshin}, \& {Martins}}]{Cuadra2008}
{Cuadra}, J., {Nayakshin}, S., \& {Martins}, F. 2008, \mnras, 383, 458, \dodoi{10.1111/j.1365-2966.2007.12573.x}

\bibitem[{{Curd} \& {Narayan}(2019)}]{CurdNarayan2019}
{Curd}, B., \& {Narayan}, R. 2019, \mnras, 483, 565, \dodoi{10.1093/mnras/sty3134}

\bibitem[{{Dai} {et~al.}(2018){Dai}, {McKinney}, {Roth}, {Ramirez-Ruiz}, \& {Miller}}]{Dai+2018}
{Dai}, L., {McKinney}, J.~C., {Roth}, N., {Ramirez-Ruiz}, E., \& {Miller}, M.~C. 2018, \apjl, 859, L20, \dodoi{10.3847/2041-8213/aab429}

\bibitem[{{David-Uraz} {et~al.}(2021){David-Uraz}, {Petit}, {Shultz}, {Fullerton}, {Erba}, {Keszthelyi}, {Seadrow}, \& {Wade}}]{David-Uraz2021}
{David-Uraz}, A., {Petit}, V., {Shultz}, M.~E., {et~al.} 2021, \mnras, 501, 2677, \dodoi{10.1093/mnras/staa3768}

\bibitem[{{Davies} {et~al.}(2012){Davies}, {Burtscher}, {Dodds-Eden}, \& {Orban de Xivry}}]{Davies2012}
{Davies}, R., {Burtscher}, L., {Dodds-Eden}, K., \& {Orban de Xivry}, G. 2012, in Journal of Physics Conference Series, Vol. 372, Journal of Physics Conference Series (IOP), 012046, \dodoi{10.1088/1742-6596/372/1/012046}

\bibitem[{{de Mink} {et~al.}(2014){de Mink}, {Sana}, {Langer}, {Izzard}, \& {Schneider}}]{deMink2014}
{de Mink}, S.~E., {Sana}, H., {Langer}, N., {Izzard}, R.~G., \& {Schneider}, F.~R.~N. 2014, \apj, 782, 7, \dodoi{10.1088/0004-637X/782/1/7}

\bibitem[{{De Villiers} {et~al.}(2003){De Villiers}, {Hawley}, \& {Krolik}}]{DeVilliers+2003}
{De Villiers}, J.-P., {Hawley}, J.~F., \& {Krolik}, J.~H. 2003, \apj, 599, 1238, \dodoi{10.1086/379509}

\bibitem[{{Donati} \& {Landstreet}(2009)}]{Donati2009}
{Donati}, J.-F., \& {Landstreet}, J.~D. 2009, \araa, 47, 333, \dodoi{10.1146/annurev-astro-082708-101833}

\bibitem[{{Donati} {et~al.}(2012){Donati}, {Gregory}, {Alencar}, {Hussain}, {Bouvier}, {Dougados}, {Jardine}, {M{\'e}nard}, \& {Romanova}}]{Donati2012}
{Donati}, J.-F., {Gregory}, S.~G., {Alencar}, S.~H.~P., {et~al.} 2012, \mnras, 425, 2948, \dodoi{10.1111/j.1365-2966.2012.21482.x}

\bibitem[{{Event Horizon Telescope Collaboration}(2021)}]{EHT_M87_Magnetic_Field_2021}
{Event Horizon Telescope Collaboration}. 2021, The Astrophysical Journal Letters, 910, L13, \dodoi{10.3847/2041-8213/abe4de}

\bibitem[{{Event Horizon Telescope Collaboration} {et~al.}(2019){Event Horizon Telescope Collaboration}, {Akiyama}, {Alberdi}, {Alef}, {Asada}, {Azulay}, {Baczko}, {Ball}, {Balokovi{\'c}}, {Barrett}, {Bintley}, {Blackburn}, {Boland}, {Bouman}, {Bower}, {Bremer}, {Brinkerink}, {Brissenden}, {Britzen}, {Broderick}, {Broguiere}, {Bronzwaer}, {Byun}, {Carlstrom}, {Chael}, {Chan}, {Chatterjee}, {Chatterjee}, {Chen}, {Chen}, {Cho}, {Christian}, {Conway}, {Cordes}, {Crew}, {Cui}, {Davelaar}, {De Laurentis}, {Deane}, {Dempsey}, {Desvignes}, {Dexter}, {Doeleman}, {Eatough}, {Falcke}, {Fish}, {Fomalont}, {Fraga-Encinas}, {Friberg}, {Fromm}, {G{\'o}mez}, {Galison}, {Gammie}, {Garc{\'\i}a}, {Gentaz}, {Georgiev}, {Goddi}, {Gold}, {Gu}, {Gurwell}, {Hada}, {Hecht}, {Hesper}, {Ho}, {Ho}, {Honma}, {Huang}, {Huang}, {Hughes}, {Ikeda}, {Inoue}, {Issaoun}, {James}, {Jannuzi}, {Janssen}, {Jeter}, {Jiang}, {Johnson}, {Jorstad}, {Jung}, {Karami}, {Karuppusamy}, {Kawashima}, {Keating}, {Kettenis}, {Kim}, {Kim}, {Kim}, {Kino},
  {Koay}, {Koch}, {Koyama}, {Kramer}, {Kramer}, {Krichbaum}, {Kuo}, {Lauer}, {Lee}, {Li}, {Li}, {Lindqvist}, {Liu}, {Liuzzo}, {Lo}, {Lobanov}, {Loinard}, {Lonsdale}, {Lu}, {MacDonald}, {Mao}, {Markoff}, {Marrone}, {Marscher}, {Mart{\'\i}-Vidal}, {Matsushita}, {Matthews}, {Medeiros}, {Menten}, {Mizuno}, {Mizuno}, {Moran}, {Moriyama}, {Moscibrodzka}, {Mul{\ensuremath{\ddot{}}}ler}, {Nagai}, {Nagar}, {Nakamura}, {Narayan}, {Narayanan}, {Natarajan}, {Neri}, {Ni}, {Noutsos}, {Okino}, {Olivares}, {Oyama}, {{\"O}zel}, {Palumbo}, {Patel}, {Pen}, {Pesce}, {Pi{\'e}tu}, {Plambeck}, {PopStefanija}, {Porth}, {Prather}, {Preciado-L{\'o}pez}, {Psaltis}, {Pu}, {Ramakrishnan}, {Rao}, {Rawlings}, {Raymond}, {Rezzolla}, {Ripperda}, {Roelofs}, {Rogers}, {Ros}, {Rose}, {Roshanineshat}, {Rottmann}, {Roy}, {Ruszczyk}, {Ryan}, {Rygl}, {S{\'a}nchez}, {S{\'a}nchez-Arguelles}, {Sasada}, {Savolainen}, {Schloerb}, {Schuster}, {Shao}, {Shen}, {Small}, {Sohn}, {SooHoo}, {Tazaki}, {Tiede}, {Tilanus}, {Titus}, {Toma}, {Torne}, {Trent},
  {Trippe}, {Tsuda}, {van Bemmel}, {van Langevelde}, {van Rossum}, {Wagner}, {Wardle}, {Weintroub}, {Wex}, {Wharton}, {Wielgus}, {Wong}, {Wu}, {Young}, {Young}, {Younsi}, \& {Yuan}}]{EHTV2019}
{Event Horizon Telescope Collaboration}, {Akiyama}, K., {Alberdi}, A., {et~al.} 2019, \apjl, 875, L5, \dodoi{10.3847/2041-8213/ab0f43}

\bibitem[{{Event Horizon Telescope Collaboration} {et~al.}(2024){Event Horizon Telescope Collaboration}, {Akiyama}, {Alberdi}, {Alef}, {Algaba}, {Anantua}, {Asada}, {Azulay}, {Bach}, {Baczko}, {Ball}, {Balokovi{\'c}}, {Bandyopadhyay}, {Barrett}, {Baub{\"o}ck}, {Benson}, {Bintley}, {Blackburn}, {Blundell}, {Bouman}, {Bower}, {Boyce}, {Bremer}, {Brinkerink}, {Brissenden}, {Britzen}, {Broderick}, {Broguiere}, {Bronzwaer}, {Bustamante}, {Byun}, {Carlstrom}, {Ceccobello}, {Chael}, {Chan}, {Chang}, {Chatterjee}, {Chatterjee}, {Chen}, {Chen}, {Cheng}, {Cho}, {Christian}, {Conroy}, {Conway}, {Cordes}, {Crawford}, {Crew}, {Cruz-Osorio}, {Cui}, {Dahale}, {Davelaar}, {De Laurentis}, {Deane}, {Dempsey}, {Desvignes}, {Dexter}, {Dhruv}, {Dihingia}, {Doeleman}, {Dougall}, {Dzib}, {Eatough}, {Emami}, {Falcke}, {Farah}, {Fish}, {Fomalont}, {Ford}, {Foschi}, {Fraga-Encinas}, {Freeman}, {Friberg}, {Fromm}, {Fuentes}, {Galison}, {Gammie}, {Garc{\'\i}a}, {Gentaz}, {Georgiev}, {Goddi}, {Gold}, {G{\'o}mez-Ruiz}, {G{\'o}mez}, {Gu},
  {Gurwell}, {Hada}, {Haggard}, {Haworth}, {Hecht}, {Hesper}, {Heumann}, {Ho}, {Ho}, {Honma}, {Huang}, {Huang}, {Hughes}, {Ikeda}, {Impellizzeri}, {Inoue}, {Issaoun}, {James}, {Jannuzi}, {Janssen}, {Jeter}, {Jiang}, {Jim{\'e}nez-Rosales}, {Johnson}, {Jorstad}, {Joshi}, {Jung}, {Karami}, {Karuppusamy}, {Kawashima}, {Keating}, {Kettenis}, {Kim}, {Kim}, {Kim}, {Kim}, {Kino}, {Koay}, {Kocherlakota}, {Kofuji}, {Koch}, {Koyama}, {Kramer}, {Kramer}, {Kramer}, {Krichbaum}, {Kuo}, {La Bella}, {Lauer}, {Lee}, {Lee}, {Leung}, {Levis}, {Li}, {Lico}, {Lindahl}, {Lindqvist}, {Lisakov}, {Liu}, {Liu}, {Liuzzo}, {Lo}, {Lobanov}, {Loinard}, {Lonsdale}, {Lowitz}, {Lu}, {MacDonald}, {Mao}, {Marchili}, {Markoff}, {Marrone}, {Marscher}, {Mart{\'\i}-Vidal}, {Matsushita}, {Matthews}, {Medeiros}, {Menten}, {Michalik}, {Mizuno}, {Mizuno}, {Moran}, {Moriyama}, {Moscibrodzka}, {Mulaudzi}, {M{\"u}ller}, {M{\"u}ller}, {Mus}, {Musoke}, {Myserlis}, {Nadolski}, {Nagai}, {Nagar}, {Nakamura}, {Narayanan}, {Natarajan}, {Nathanail}, {Fuentes},
  {Neilsen}, {Neri}, {Ni}, {Noutsos}, {Nowak}, {Oh}, {Okino}, {Olivares}, {Ortiz-Le{\'o}n}, {Oyama}, {{\"O}zel}, {Palumbo}, {Paraschos}, {Park}, {Parsons}, {Patel}, \& {Pen}}]{EHT8}
---. 2024, \apjl, 964, L26, \dodoi{10.3847/2041-8213/ad2df1}

\bibitem[{{Ferrario} {et~al.}(2015){Ferrario}, {de Martino}, \& {G{\"a}nsicke}}]{Ferrario2015}
{Ferrario}, L., {de Martino}, D., \& {G{\"a}nsicke}, B.~T. 2015, \ssr, 191, 111, \dodoi{10.1007/s11214-015-0152-0}

\bibitem[{{Galishnikova} {et~al.}(2025){Galishnikova}, {Philippov}, {Quataert}, {Chatterjee}, \& {Liska}}]{Galishnikova2025}
{Galishnikova}, A., {Philippov}, A., {Quataert}, E., {Chatterjee}, K., \& {Liska}, M. 2025, \apj, 978, 148, \dodoi{10.3847/1538-4357/ad9926}

\bibitem[{{Gezari}(2021)}]{Gezari2021}
{Gezari}, S. 2021, \araa, 59, 21, \dodoi{10.1146/annurev-astro-111720-030029}

\bibitem[{{Gourgouliatos} \& {Cumming}(2014)}]{Gourgouliatos2014}
{Gourgouliatos}, K.~N., \& {Cumming}, A. 2014, \mnras, 438, 1618, \dodoi{10.1093/mnras/stt2300}

\bibitem[{{Guillochon} \& {McCourt}(2017)}]{GuillochonMcCourt2017}
{Guillochon}, J., \& {McCourt}, M. 2017, \apjl, 834, L19, \dodoi{10.3847/2041-8213/834/2/L19}

\bibitem[{{Guolo} {et~al.}(2024){Guolo}, {Gezari}, {Yao}, {van Velzen}, {Hammerstein}, {Cenko}, \& {Tokayer}}]{Guolo+2024}
{Guolo}, M., {Gezari}, S., {Yao}, Y., {et~al.} 2024, \apj, 966, 160, \dodoi{10.3847/1538-4357/ad2f9f}

\bibitem[{{Hawley} \& {Krolik}(2001)}]{HK2001}
{Hawley}, J.~F., \& {Krolik}, J.~H. 2001, \apj, 548, 348, \dodoi{10.1086/318678}

\bibitem[{{Hawley} \& {Krolik}(2006)}]{HK2006}
---. 2006, \apj, 641, 103, \dodoi{10.1086/500385}

\bibitem[{{Hobson} {et~al.}(2007){Hobson}, {Efstathiou}, {Lasenby}, \& {Ford}}]{Hobson+2007}
{Hobson}, M.~P., {Efstathiou}, G., {Lasenby}, A.~N., \& {Ford}, L.~H. 2007, Physics Today, 60, 62, \dodoi{10.1063/1.2718760}

\bibitem[{{Huang} {et~al.}(2023){Huang}, {Jiang}, {Feng}, {Davis}, {Stone}, \& {Middleton}}]{Huang+2023b}
{Huang}, J., {Jiang}, Y.-F., {Feng}, H., {et~al.} 2023, \apj, 945, 57, \dodoi{10.3847/1538-4357/acb6fc}

\bibitem[{{Jiang} {et~al.}(2019){Jiang}, {Blaes}, {Stone}, \& {Davis}}]{Jiang2019}
{Jiang}, Y.-F., {Blaes}, O., {Stone}, J.~M., \& {Davis}, S.~W. 2019, \apj, 885, 144, \dodoi{10.3847/1538-4357/ab4a00}

\bibitem[{{Johns-Krull}(2007)}]{JohnsKrull2007}
{Johns-Krull}, C.~M. 2007, \apj, 664, 975, \dodoi{10.1086/519017}

\bibitem[{{Kelley} {et~al.}(2014){Kelley}, {Tchekhovskoy}, \& {Narayan}}]{KelleyTchekhovskoy2014}
{Kelley}, L.~Z., {Tchekhovskoy}, A., \& {Narayan}, R. 2014, \mnras, 445, 3919, \dodoi{10.1093/mnras/stu2041}

\bibitem[{{Krolik} {et~al.}(2025){Krolik}, {Piran}, \& {Ryu}}]{Krolik+2025}
{Krolik}, J., {Piran}, T., \& {Ryu}, T. 2025, \apj, 988, 220, \dodoi{10.3847/1538-4357/ade797}

\bibitem[{{Krolik} \& {Piran}(2012)}]{KrolikPiran2012}
{Krolik}, J.~H., \& {Piran}, T. 2012, \apj, 749, 92, \dodoi{10.1088/0004-637X/749/1/92}

\bibitem[{{Kroupa}(2001)}]{Kroupa2001}
{Kroupa}, P. 2001, \mnras, 322, 231, \dodoi{10.1046/j.1365-8711.2001.04022.x}

\bibitem[{{Kwan} {et~al.}(2023){Kwan}, {Dai}, \& {Tchekhovskoy}}]{Kwan2023}
{Kwan}, T.~M., {Dai}, L., \& {Tchekhovskoy}, A. 2023, \apjl, 946, L42, \dodoi{10.3847/2041-8213/acc334}

\bibitem[{{Lalakos} {et~al.}(2024){Lalakos}, {Tchekhovskoy}, {Bromberg}, {Gottlieb}, {Jacquemin-Ide}, {Liska}, \& {Zhang}}]{Lalakos2024}
{Lalakos}, A., {Tchekhovskoy}, A., {Bromberg}, O., {et~al.} 2024, \apj, 964, 79, \dodoi{10.3847/1538-4357/ad0974}

\bibitem[{{Ligni{\`e}res} {et~al.}(2014){Ligni{\`e}res}, {Petit}, {Auri{\`e}re}, {Wade}, \& {B{\"o}hm}}]{Lignieres2014}
{Ligni{\`e}res}, F., {Petit}, P., {Auri{\`e}re}, M., {Wade}, G.~A., \& {B{\"o}hm}, T. 2014, in IAU Symposium, Vol. 302, Magnetic Fields throughout Stellar Evolution, ed. P.~{Petit}, M.~{Jardine}, \& H.~C. {Spruit}, 338--347, \dodoi{10.1017/S1743921314002440}

\bibitem[{{Mandel} \& {Levin}(2015)}]{Mandel2015}
{Mandel}, I., \& {Levin}, Y. 2015, \apjl, 805, L4, \dodoi{10.1088/2041-8205/805/1/L4}

\bibitem[{{McKinney}(2005)}]{McKinney2005}
{McKinney}, J.~C. 2005, \apjl, 630, L5, \dodoi{10.1086/468184}

\bibitem[{{McKinney} \& {Gammie}(2004)}]{McK+Gammie2004}
{McKinney}, J.~C., \& {Gammie}, C.~F. 2004, \apj, 611, 977, \dodoi{10.1086/422244}

\bibitem[{{Parker}(1955)}]{Parker1955}
{Parker}, E.~N. 1955, \apj, 122, 293, \dodoi{10.1086/146087}

\bibitem[{{Pasham} {et~al.}(2023){Pasham}, {Lucchini}, {Laskar}, {Gompertz}, {Srivastav}, {Nicholl}, {Smartt}, {Miller-Jones}, {Alexander}, {Fender}, {Smith}, {Fulton}, {Dewangan}, {Gendreau}, {Coughlin}, {Rhodes}, {Horesh}, {van Velzen}, {Sfaradi}, {Guolo}, {Castro Segura}, {Aamer}, {Anderson}, {Arcavi}, {Brennan}, {Chambers}, {Charalampopoulos}, {Chen}, {Clocchiatti}, {de Boer}, {Dennefeld}, {Ferrara}, {Galbany}, {Gao}, {Gillanders}, {Goodwin}, {Gromadzki}, {Huber}, {Jonker}, {Joshi}, {Kara}, {Killestein}, {Kosec}, {Kocevski}, {Leloudas}, {Lin}, {Margutti}, {Mattila}, {Moore}, {M{\"u}ller-Bravo}, {Ngeow}, {Oates}, {Onori}, {Pan}, {Perez-Torres}, {Rani}, {Remillard}, {Ridley}, {Schulze}, {Sheng}, {Shingles}, {Smith}, {Steiner}, {Wainscoat}, {Wevers}, \& {Yang}}]{Pasham+2023}
{Pasham}, D.~R., {Lucchini}, M., {Laskar}, T., {et~al.} 2023, Nature Astronomy, 7, 88, \dodoi{10.1038/s41550-022-01820-x}

\bibitem[{{Petit} {et~al.}(2008){Petit}, {Dintrans}, {Solanki}, {Donati}, {Auri{\`e}re}, {Ligni{\`e}res}, {Morin}, {Paletou}, {Ramirez Velez}, {Catala}, \& {Fares}}]{Petit2008}
{Petit}, P., {Dintrans}, B., {Solanki}, S.~K., {et~al.} 2008, \mnras, 388, 80, \dodoi{10.1111/j.1365-2966.2008.13411.x}

\bibitem[{{Piran} {et~al.}(2015{\natexlab{a}}){Piran}, {S{\k{a}}dowski}, \& {Tchekhovskoy}}]{Piran2015SadTch}
{Piran}, T., {S{\k{a}}dowski}, A., \& {Tchekhovskoy}, A. 2015{\natexlab{a}}, \mnras, 453, 157, \dodoi{10.1093/mnras/stv1547}

\bibitem[{{Piran} {et~al.}(2015{\natexlab{b}}){Piran}, {Svirski}, {Krolik}, {Cheng}, \& {Shiokawa}}]{Piran+2015}
{Piran}, T., {Svirski}, G., {Krolik}, J., {Cheng}, R.~M., \& {Shiokawa}, H. 2015{\natexlab{b}}, \apj, 806, 164, \dodoi{10.1088/0004-637X/806/2/164}

\bibitem[{{Price} {et~al.}(2024){Price}, {Liptai}, {Mandel}, {Shepherd}, {Lodato}, \& {Levin}}]{Price+2024}
{Price}, D.~J., {Liptai}, D., {Mandel}, I., {et~al.} 2024, arXiv e-prints, arXiv:2404.09381, \dodoi{10.48550/arXiv.2404.09381}

\bibitem[{{Rees}(1988)}]{Rees1988}
{Rees}, M.~J. 1988, \nat, 333, 523, \dodoi{10.1038/333523a0}

\bibitem[{{Ressler} {et~al.}(2021){Ressler}, {Quataert}, {White}, \& {Blaes}}]{Ressler2021}
{Ressler}, S.~M., {Quataert}, E., {White}, C.~J., \& {Blaes}, O. 2021, \mnras, 504, 6076, \dodoi{10.1093/mnras/stab311}

\bibitem[{{Romanyuk} \& {Moiseeva}(2025)}]{Romanyuk2025}
{Romanyuk}, I.~I., \& {Moiseeva}, A.~V. 2025, Astrophysical Bulletin, 80, 468, \dodoi{10.1134/S199034132560036X}

\bibitem[{{Ryu} {et~al.}(2023{\natexlab{a}}){Ryu}, {Krolik}, \& {Piran}}]{Ryu+2023}
{Ryu}, T., {Krolik}, J., \& {Piran}, T. 2023{\natexlab{a}}, \apjl, 946, L33, \dodoi{10.3847/2041-8213/acc390}

\bibitem[{{Ryu} {et~al.}(2020{\natexlab{a}}){Ryu}, {Krolik}, {Piran}, \& {Noble}}]{Ryu+2020b}
{Ryu}, T., {Krolik}, J., {Piran}, T., \& {Noble}, S.~C. 2020{\natexlab{a}}, \apj, 904, 99, \dodoi{10.3847/1538-4357/abb3cd}

\bibitem[{{Ryu} {et~al.}(2020{\natexlab{b}}){Ryu}, {Krolik}, {Piran}, \& {Noble}}]{Ryu+2020a}
---. 2020{\natexlab{b}}, \apj, 904, 98, \dodoi{10.3847/1538-4357/abb3cf}

\bibitem[{{Ryu} {et~al.}(2023{\natexlab{b}}){Ryu}, {Krolik}, {Piran}, {Noble}, \& {Avara}}]{ShocksPowerTDEs2023}
{Ryu}, T., {Krolik}, J., {Piran}, T., {Noble}, S.~C., \& {Avara}, M. 2023{\natexlab{b}}, \apj, 957, 12, \dodoi{10.3847/1538-4357/acf5de}

\bibitem[{{Schneider} {et~al.}(2019){Schneider}, {Ohlmann}, {Podsiadlowski}, {R{\"o}pke}, {Balbus}, {Pakmor}, \& {Springel}}]{Schneider2019}
{Schneider}, F. R.~N., {Ohlmann}, S.~T., {Podsiadlowski}, P., {et~al.} 2019, \nat, 574, 211, \dodoi{10.1038/s41586-019-1621-5}

\bibitem[{{Sen} {et~al.}(2025){Sen}, {Maity}, \& {Das}}]{Sen2025}
{Sen}, G., {Maity}, D., \& {Das}, S. 2025, \prd, 112, 083047, \dodoi{10.1103/v8mp-hks1}

\bibitem[{{Sfaradi} {et~al.}(2024){Sfaradi}, {Beniamini}, {Horesh}, {Piran}, {Bright}, {Rhodes}, {Williams}, {Fender}, {Leung}, {Murphy}, \& et~al.}]{Sfaradi+2024}
{Sfaradi}, I., {Beniamini}, P., {Horesh}, A., {et~al.} 2024, \mnras, 527, 7672, \dodoi{10.1093/mnras/stad3717}

\bibitem[{Shakura \& Sunyaev(1973)}]{Shakura1973}
Shakura, N.~I., \& Sunyaev, R.~A. 1973, Astronomy and Astrophysics, 24, 337.
\newblock \url{https://ui.adsabs.harvard.edu/abs/1973A\%26A....24..337S}

\bibitem[{{Shenar} {et~al.}(2023){Shenar}, {Wade}, {Marchant}, {Bagnulo}, {Bodensteiner}, {Bowman}, {Gilkis}, {Langer}, {Nicolas-Chen{\'e}}, {Oskinova}, {Van Reeth}, {Sana}, {St-Louis}, {de Oliveira}, {Todt}, \& {Toonen}}]{Shenar2023}
{Shenar}, T., {Wade}, G.~A., {Marchant}, P., {et~al.} 2023, Science, 381, 761, \dodoi{10.1126/science.ade3293}

\bibitem[{{Sikora} {et~al.}(2019){Sikora}, {Wade}, {Power}, \& {Neiner}}]{Sikora2019}
{Sikora}, J., {Wade}, G.~A., {Power}, J., \& {Neiner}, C. 2019, \mnras, 483, 2300, \dodoi{10.1093/mnras/sty3105}

\bibitem[{{S{\k{a}}dowski} {et~al.}(2014){S{\k{a}}dowski}, {Narayan}, {McKinney}, \& {Tchekhovskoy}}]{SadowskiNarayan+2014}
{S{\k{a}}dowski}, A., {Narayan}, R., {McKinney}, J.~C., \& {Tchekhovskoy}, A. 2014, \mnras, 439, 503, \dodoi{10.1093/mnras/stt2479}

\bibitem[{{Steinberg} \& {Stone}(2024)}]{SteinbergStone2024}
{Steinberg}, E., \& {Stone}, N.~C. 2024, \nat, 625, 463, \dodoi{10.1038/s41586-023-06875-y}

\bibitem[{{Stenflo}(2013)}]{Stenflo2013}
{Stenflo}, J.~O. 2013, \aapr, 21, 66, \dodoi{10.1007/s00159-013-0066-3}

\bibitem[{{Takata} {et~al.}(2026){Takata}, {Murphy}, {Kurtz}, {Saio}, \& {Shibahashi}}]{Takata2026}
{Takata}, M., {Murphy}, S.~J., {Kurtz}, D.~W., {Saio}, H., \& {Shibahashi}, H. 2026, \mnras, 545, staf2153, \dodoi{10.1093/mnras/staf2153}

\bibitem[{{Tchekhovskoy} {et~al.}(2014){Tchekhovskoy}, {Metzger}, {Giannios}, \& {Kelley}}]{TchekhovskoyMetzger2014}
{Tchekhovskoy}, A., {Metzger}, B.~D., {Giannios}, D., \& {Kelley}, L.~Z. 2014, \mnras, 437, 2744, \dodoi{10.1093/mnras/stt2085}

\bibitem[{{Tchekhovskoy} {et~al.}(2011){Tchekhovskoy}, {Narayan}, \& {McKinney}}]{Tchekhovskoy2011}
{Tchekhovskoy}, A., {Narayan}, R., \& {McKinney}, J.~C. 2011, \mnras, 418, L79, \dodoi{10.1111/j.1745-3933.2011.01147.x}

\bibitem[{{van Velzen} {et~al.}(2013){van Velzen}, {Frail}, {K{\"o}rding}, \& {Falcke}}]{vanVelzen2013}
{van Velzen}, S., {Frail}, D.~A., {K{\"o}rding}, E., \& {Falcke}, H. 2013, \aap, 552, A5, \dodoi{10.1051/0004-6361/201220426}

\bibitem[{{Wade} {et~al.}(2012){Wade}, {Ma{\'\i}z Apell{\'a}niz}, {Martins}, {Petit}, {Grunhut}, {Walborn}, {Barb{\'a}}, {Gagn{\'e}}, {Garc{\'\i}a-Melendo}, {Jose}, {Moffat}, {Naz{\'e}}, {Neiner}, {Pellerin}, {Penad{\'e}s Ordaz}, {Shultz}, {Sim{\'o}n-D{\'\i}az}, \& {Sota}}]{Wade2012_2}
{Wade}, G.~A., {Ma{\'\i}z Apell{\'a}niz}, J., {Martins}, F., {et~al.} 2012, \mnras, 425, 1278, \dodoi{10.1111/j.1365-2966.2012.21523.x}

\bibitem[{{Zauderer} {et~al.}(2013){Zauderer}, {Berger}, {Margutti}, {Pooley}, {Sari}, {Soderberg}, {Brunthaler}, \& {Bietenholz}}]{Zauderer2013}
{Zauderer}, B.~A., {Berger}, E., {Margutti}, R., {et~al.} 2013, \apj, 767, 152, \dodoi{10.1088/0004-637X/767/2/152}

\bibitem[{{Zhang} {et~al.}(2026){Zhang}, {Stone}, {Davis}, {Jiang}, {Mullen}, \& {White}}]{Zhang+2026}
{Zhang}, L., {Stone}, J.~M., {Davis}, S.~W., {et~al.} 2026, arXiv e-prints, arXiv:2603.05588, \dodoi{10.48550/arXiv.2603.05588}

\bibitem[{{Zhu} {et~al.}(2015){Zhu}, {Pakmor}, {van Kerkwijk}, \& {Chang}}]{Zhu2015}
{Zhu}, C., {Pakmor}, R., {van Kerkwijk}, M.~H., \& {Chang}, P. 2015, \apjl, 806, L1, \dodoi{10.1088/2041-8205/806/1/L1}

\end{thebibliography}

\end{document}